\documentclass[reprint,twocolumn,pre,showpacs,amsmath,amssymb,aps]{revtex4-1}

\usepackage{amsmath}
\usepackage{natbib}	
\usepackage{graphicx,color,rotating}
\usepackage[latin1]{inputenc}
\usepackage{textcomp}
\usepackage{dcolumn}% Align table columns on decimal point
\usepackage{bm}     % bold math
\usepackage{upgreek}
\usepackage{subdepth}  			% Added by JTK (aligns sub-indices)
\usepackage{multirow}
\usepackage[latin1]{inputenc}
\usepackage{hyperref}						% Package for hyperreferences
\usepackage{xcolor}						% Allows for coloring of text
\hypersetup{							% Setup of the Hyperref package
    colorlinks,
    linkcolor={blue!80!black},
    citecolor={blue!80!black},
    urlcolor={blue!80!black}
}
\newcommand{\mr}[1]{\ensuremath{\mathrm{#1}}}
\renewcommand{\vec}[1]{\bm{#1}}
\newcommand{\ee}{\mathrm{e}}
\newcommand{\ii}{\mathrm{i}}

\newcommand{\avr}[1]{\big\langle #1 \big\rangle}
\newcommand{\abs}[1]{\big|{#1}\big|}
\newcommand{\sign}{\mr{sign}}

\DeclareMathOperator{\im}{Im}

\newcommand{\xiti}{\tilde{\xi}}

\newcommand{\pp}{\partial}

\newcommand{\nablabf}{\boldsymbol{\nabla}}

\newcommand{\Lstar}{L^\star}
\newcommand{\lp}{l'\hspace{-1.5pt}}
\renewcommand{\mp}{m\hspace{-1pt}'\hspace{-1.5pt}}
\newcommand{\np}{n\hspace{-1pt}'\hspace{-1.5pt}}
\newcommand{\lmnp}{{\lp\mp\np}}
\newcommand{\lmp}{\lp\mp}

\newcommand{\curl}{\nablabf\times}
\newcommand{\rot}{\curl}

\newcommand{\ie}{\textit{i.e.}}

\newcommand{\etal}{\textit{et~al.}}

\newcommand{\qmarks}[1]{``{#1}"}

\newcommand*{\plimsoll}{{\ensuremath{-\kern-4pt{\ominus}\kern-4pt-}}}

\newcommand{\cfl}{c_\mr{fl}}

\newcommand{\D}{\mathcal{D}}
\newcommand{\Dx}{\mathcal{D}_x}

\newcommand{\Dz}{\mathcal{D}_z}

\newcommand{\dd}{\mr{d}}
\newcommand{\ddd}{\vec{d}}

\newcommand{\eee}{\vec{e}}
\newcommand{\een}{\vec{e}}

\newcommand{\fff}{\vec{f}}

\newcommand{\fffac}{\vec{f}_\mathrm{ac}}

\newcommand{\kc}{k_\mathrm{c}}

\newcommand{\kti}{\tilde{k}}
\newcommand{\kx}{k_x}
\newcommand{\ky}{k_y}
\newcommand{\kz}{k_z}
\newcommand{\Kx}{K_x}
\newcommand{\Ky}{K_y}
\newcommand{\Kz}{K_z}

\newcommand{\Deltati}{\tilde{\Delta}}

\newcommand{\kxnx}{k_x^{\nx}}
\newcommand{\kyny}{k_y^{\ny}}
\newcommand{\kznz}{k_z^{\nz}}
\newcommand{\Kxnx}{K_x^{\nx}}
\newcommand{\Kyny}{K_y^{\ny}}
\newcommand{\Kznz}{K_z^{\nz}}

\newcommand{\ks}{k_\mathrm{s}}

\newcommand{\nnn}{\vec{n}}

\newcommand{\rrr}{\vec{r}}

\newcommand{\SSS}{\vec{S}}
\newcommand{\SSSac}{\vec{S}_\mathrm{ac}}

\newcommand{\uuu}{\vec{u}}

\newcommand{\vvv}{\vec{v}}

\newcommand{\Uhat}{\hat{U}}

\newcommand{\zerovec}{\boldsymbol{0}}

\newcommand{\calF}{\mathcal{F}}

\newcommand{\calKKK}{\vec{\mathcal{K}}}

\newcommand{\calLLL}{\vec{\mathcal{L}}}

\newcommand{\deltak}{\eps}
\newcommand{\deltakxnx}{\deltak_x^l}

\newcommand{\deltakOlmn}{\deltak_0^{lmn}}

\newcommand{\Ekin}{E_\mathrm{kin}}

\newcommand{\Epot}{E_\mathrm{pot}}

\newcommand{\Eac}{E_\mathrm{ac}}

\newcommand{\kapfl}{\kappa_\mathrm{fl}}

\newcommand{\KO}{K_0}

\newcommand{\Lx}{L_x}
\newcommand{\Ly}{L_y}
\newcommand{\Lz}{L_z}
\newcommand{\Lxnx}{L_x^{{\nx}^{}_{}}}
\newcommand{\Lyny}{L_y^{{\ny}^{}_{}}}
\newcommand{\Lznz}{L_z^{{\nz}^{}_{}}}
\newcommand{\Lac}{\bm{\mathcal{L}}_\mathrm{ac}}

\newcommand{\alphaxnx}{{\alpha_x^{\nx}}}
\newcommand{\alphayny}{{\alpha_y^{\ny}}}
\newcommand{\alphaznz}{{\alpha_z^{\nz}}}

\newcommand{\eps}{\epsilon}

\newcommand{\etafl}{\eta_\mr{fl}}

\newcommand{\etaB}{\eta^\mathrm{b}}
\newcommand{\etaBfl}{\eta^\mathrm{b}_\mathrm{fl}}

\newcommand{\Gambl}{\Gamma_\mathrm{bl}}
\newcommand{\Gamfl}{\Gamma_\mathrm{fl}}

\newcommand{\phiF}{\phi_F}
\newcommand{\phiact}{\phi_\mr{act}}

\newcommand{\pfl}{p_\mathrm{fl}}

\newcommand{\rhofl}{\rho_\mathrm{fl}}

\newcommand{\kO}{k_{0}}

\newcommand{\nx}{l}
\newcommand{\ny}{m}
\newcommand{\nz}{n}

\newcommand{\pa}{P_1}

\newcommand{\phitot}{\phi_\mathrm{tot}}

\newcommand{\SIum}{\upmu\textrm{m}}

\newcommand{\SIMHz}{\textrm{MHz}}

\newcommand{\SIkg}{\textrm{kg}}

\newcommand{\SIm}{\textrm{m}}

\newcommand{\SImm}{\textrm{mm}}

\newcommand{\SInm}{\textrm{nm}}
\newcommand{\SIN}{\textrm{N}}

\newcommand{\SIMPa}{\textrm{MPa}}

\newcommand{\SIs}{\textrm{s}}

\newcommand{\nn}{\nonumber}
\newcommand{\beq}[1]{\begin{equation} \eqlab{#1}}
\newcommand{\eeq}{\end{equation}}
\newcommand{\bsub}{\begin{subequations}}
\newcommand{\esub}{\end{subequations}}
\def\bal#1\eal{\begin{align}#1\end{align}}
\def\balat#1#2\ealat{\begin{alignat}{#1} #2 \end{alignat}}
\def\bsublab#1#2\esublab{\bsub \eqlab{#1} #2 \esub}
\def\bsubal#1#2\esubal{\bsublab{#1}\begin{align}#2\end{align} \esublab}% begin/end align with a,b,c-equation labels
\def\bsubalat#1#2#3\esubalat{\bsublab{#1} \begin{alignat}{#2} #3 \end{alignat} \esublab}

\newcommand{\eqlab}[1]{\label{eq:#1}}
\renewcommand{\eqref}[1]{Eq.~(\ref{eq:#1})}
\newcommand{\eqnoref}[1]{(\ref{eq:#1})}
\newcommand{\eqrefnoEq}[1]{(\ref{eq:#1})}
\newcommand{\eqsref}[2]{Eqs.~(\ref{eq:#1}) and~(\ref{eq:#2})}

\newcommand{\eqssref}[3]{Eqs.~(\ref{eq:#1}), (\ref{eq:#2}) and~(\ref{eq:#3})}

\newcommand{\figref}[1]{Fig.~\ref{fig:#1}}
\newcommand{\fignoref}[1]{\ref{fig:#1}}

\newcommand{\figlab}[1]{\label{fig:#1}}
\newcommand{\appref}[1]{Appendix~\ref{sec:#1}}

\newcommand{\secref}[1]{Section~\ref{sec:#1}}
\newcommand{\secrefnoSec}[1]{\ref{sec:#1}}
\newcommand{\secsref}[2]{Sections~\ref{sec:#1} and~\ref{sec:#2}}

\newcommand{\seclab}[1]{\label{sec:#1}}
\newcommand{\tabref}[1]{Table~\ref{tab:#1}}
\newcommand{\tabsref}[2]{Tables~\ref{tab:#1} and~\ref{tab:#2}}
\newcommand{\tablab}[1]{\label{tab:#1}}

\newcommand{\ord}[1]{\mathcal{O}({#1})}
\newcommand{\eiot}{\ee^{-\ii\omega t}}

\newcommand{\grad}{\boldsymbol{\nabla}}

\renewcommand{\div}{\nablabf\!\cdot}

\newcommand{\lap}{\nabla^2}

\renewcommand{\Re}{\mathrm{Re}}

\begin{document}
%\preprint{Preprint identifier}

\title{Bulk-driven acoustic streaming at resonance in closed microcavities}

\author{Jacob S. Bach}
\email{jasoba@fysik.dtu.dk}
\affiliation{Department of Physics, Technical University of Denmark,\\ DTU Physics Building 309, DK-2800 Kongens Lyngby, Denmark}

\author{Henrik Bruus}
\email{bruus@fysik.dtu.dk}
\affiliation{Department of Physics, Technical University of Denmark,\\
DTU Physics Building 309, DK-2800 Kongens Lyngby, Denmark}

\date{22 May 2019}

\begin{abstract}
Bulk-driven acoustic (Eckart) streaming is the steady flow resulting from the time-averaged acoustic energy flux density in the bulk of a viscous fluid. In simple cases, like the one-dimensional single standing-wave resonance, this energy flux is negligible, and therefore the bulk-driven streaming is often ignored relative to the boundary-driven (Rayleigh) streaming in the analysis of resonating acoustofluidic devices with length scales comparable to the acoustic wavelength. However, in closed acoustic microcavities with viscous dissipation, two overlapping resonances may be excited at the same frequency as a double mode. In contrast to single modes, the double modes can support a steady rotating acoustic energy flux density
and thus a corresponding rotating bulk-driven acoustic streaming. We derive analytical solutions for the double modes in a rectangular box-shaped cavity including the viscous boundary layers, and use them to map out possible rotating patterns of bulk-driven acoustic streaming. Remarkably, the rotating bulk-driven streaming may be excited by a non-rotating actuation, and we determine the optimal geometry that maximizes this excitation. In the optimal geometry, we finally simulate a horizontal 2-by-2, 4-by-4, and 6-by-6 streaming-roll pattern in a shallow square cavity. We find that the high-frequency 6-by-6 streaming-roll pattern is dominated by the bulk-driven streaming as opposed to the low-frequency 2-by-2 streaming pattern, which is dominated by the boundary-driven streaming.
\end{abstract}

% \pacs{43.25.Qp, 43.20.Fn, 43.20.+g, 47.35.Rs}

% 43 Acoustics
%   43.20.+g: General linear acoustics
%   43.20.Fn: Scattering of acoustic waves
%   43.20.Ks: Standing waves, resonance, normal modes
%	47.35.Rs: Sound waves in fluids
%   43.25.Gf: Standing waves; resonance
%   43.25.-x: Nonlinear acoustics
%	43.25.Nm: Acoustic streaming
%   43.25.Qp: Radiation pressure
%   43.35.Ty: Other physical effects of sound
%   43.20.Bi: Mathematical theory of wave propagation
%	43.80.-n, 43.80.+p: Ultrasound application to biology
% 47 Fluid dynamics
%   47.15.-x: laminar
%	47.35.Rs: Sound waves in fluids

%\keywords{Suggested keywords} Use showkeys class to display keywords

\maketitle

% Main text

\section{Introduction}
\seclab{intro}

Acoustophoresis, particle migration by ultrasound in a microfluidic setting, is an increasingly popular method for non-contact and label-free handling of microparticles suspended in a fluid. Examples include acoustic separation \cite{Petersson2007, Manneberg2008, Ding2014}, trapping \cite{Hammarstrom2012, Evander2015}, and tweezing \cite{Collins2016, Lim2016, Baresch2016}, as well as enrichment of cancer cells \cite{Augustsson2012, Antfolk2015} and bacteria \cite{Antfolk2014, Li2017}, and size-independent sorting of cells \cite{Augustsson2016}.

The acoustic field induces the acoustic radiation force that acts on suspended particles and scales with the particle volume \cite{King1934, Yosioka1955, Doinikov1997, Settnes2012, Karlsen2015}, as well as  the acoustic streaming in the fluid \cite{LordRayleigh1884, Schlichting1932, Westervelt1953, Nyborg1958} that gives rise to a viscous Stokes drag force on the particles which scale with the linear size of the particles \cite{Muller2013}. Consequently, the streaming-induced drag force is the dominating acoustic force for particles smaller than a critical size. For example, for an aqueous suspension of spherical polystyrene particles in a 1-MHz ultrasound field, the critical radius has been determined to be around $2~\SIum$ \cite{Muller2012, Barnkob2012a}. To extend the application of acoustofluidics into the regime of sub-micrometer particles, such as bacteria, viruses and exosomes, a deep understanding of acoustic streaming is therefore important.

The first theoretical analysis of acoustic streaming dates back to Lord Rayleigh \cite{LordRayleigh1884}, who investigated the streaming generated by the viscous boundary layers close to the walls in planar and cylindrical systems. This analysis was later generalized by Nyborg  \cite{Nyborg1958}, who formulated a slip velocity condition for the acoustic streaming outside the viscous boundary layers at weakly curved boundaries oscillating in the normal direction. Further extensions to this slip-velocity theory involves curvilinear corrections for curved boundary \cite{Lee1989}, general motion of a flat boundary \cite{Vanneste2011}, and general motion of a curved boundary \cite{Bach2018}. This kind of streaming is caused by the velocity mismatch between the boundary and the acoustic wave in the bulk, and we refer to it as boundary-driven acoustic streaming.

Acoustic streaming can also be generated by attenuation of the acoustic wave in the bulk as shown by Eckart \cite{Eckart1948}, who established the basic theory and investigated the steady flow caused by a sound beam. This kind of streaming is generated by a force, which is proportional to the acoustic energy flux density (or intensity) such that, say, a sound beam generates acoustic streaming in its direction of propagation. We refer to this kind of streaming as bulk-driven acoustic streaming.

While bulk-driven streaming is known to have a clear effect in systems much larger than the acoustic wavelength \cite{Riaud2017a}, it is often neglected in studies of resonating devices with length scales comparable with the acoustic wavelength. However, the bulk-driven streaming was taken into account implicitly by Antfolk \etal\ in their study of a long straight microchannel with a nearly-square cross section~\cite{Antfolk2014}, as they made a direct numerical simulation of the acoustofluidics in the system. They showed numerically that the experimentally observed single-vortex streaming pattern, which remarkably appeared in their microchannel instead of the usual quadrupolar Rayleigh-vortex pattern, could be explained if two orthogonal acoustic resonances with nearly identical resonance frequencies were excited with a quarter-period phase difference at the same frequency.

Such orthogonal phase-shifted acoustic waves have been used in other experiments to generate an acoustic radiation torque acting on suspended particles \cite{Busse1981, Tran2012, Dual2012d, Lee1989,Courtney2013,Mao2017a}. Also travelling Bessel beams \cite{Baresh2018,Marston2006} as well as arrays of holographic acoustic elements \cite{Marzo2015} have been used for this purpose, but not for generating specific streaming patterns.

Inspired by the work of Antfolk \etal~\cite{Antfolk2014}, we study in this paper bulk-driven acoustic streaming at resonance in rectangular-box cavities with acoustically hard walls and side lengths comparable to the acoustic wavelength. The paper is organized as follows: In \secref{GovEqu} we present the governing equations for the acoustic pressure and the acoustic streaming at resonance, where the viscous boundary layers are taken into account using our recent boundary-layer analysis \cite{Bach2018}. We apply these equations in \secref{ModeTheory} to derive analytical solutions for the acoustic single-mode pressure resonances in the rectangular cavity. We also introduce the so-called double-mode resonances, where two overlapping single modes are excited simultaneously at the same frequency. In \secref{StrongBodyForce} we establish the so-called overlap and phase conditions for double-mode resonances that lead to a strong acoustic body force, which causes the bulk-driven streaming. A main result is presented in \secref{NonRotActuation}, where we show how a weak symmetry breaking of a perfect square cavity can lead to a rotating body force even for a non-rotating actuation. We validate our theory in \secref{3DSimulation} by direct numerical simulation in the case of a shallow nearly-square cavity similar to the geometry studied experimentally by Hags\"ater \etal~\cite{Hagsater2007}.
We predict that the bulk-driven streaming in closed microcavities is enhanced by increasing the frequency and the bulk viscosity, just as is the case for Eckart streaming in open systems \cite{Eckart1948}. We propose to test this prediction by replacing water with pyridine in the microcavity. Finally, we discuss our results in \secref{Discussion} and present our conclusions in \secref{Conclusion}.

\section{Governing equations}
\seclab{GovEqu}

The time-harmonic acoustic fields in the fluid domain $\Omega$ are induced by the time-harmonic displacement field $\uuu(\rrr,t)$ of the wall at the boundary $\pp\Omega$ of the domain,
\bal
\uuu(\rrr,t)=\Re\big[\uuu_1(\rrr) \eiot\big], \qquad \rrr\in \pp\Omega,
\eal
where $\rrr$ is position, $t$ is time, $\omega=2\pi f$ is the angular driving frequency, and \qmarks{Re} is the real part of complex-valued fields. In the quiescent and homogeneous fluid of density $\rhofl$ and ambient pressure $\pfl$, we apply standard perturbation theory to describe the fluid velocity $\vvv(\rrr,t)$, pressure $p(\rrr,t)$, and density $\rho(\rrr,t)$,
 \bsubalat{pert_eqs}{3}
 \eqlab{pPert}
 p(\rrr,t) 	  &= \pfl     && +\Re\big[p_1(\rrr) \eiot\big]    &&+ p_2(\rrr), \\
 \eqlab{vPert}
 \vvv(\rrr,t) &= \zerovec && +\Re\big[\vvv_1(\rrr) \eiot\big] &&+ \vvv_2(\rrr),\\
 \eqlab{rhoPert}
 \rho(\rrr,t) &= \rhofl   && +\Re\big[\rho_1(\rrr) \eiot\big] &&+ \rho_2(\rrr),
 \esubalat
where the subscript \qmarks{1} denotes the first-order, complex-valued, time-harmonic acoustic fields and \qmarks{2} the second-order, real-valued, time-averaged steady fields. We do not compute the oscillating second-order fields containing $\ee^{\pm\ii2\omega t}$, as they are typically not observed.

\subsection{Pressure acoustics with boundary layers}

The complex-valued acoustic pressure $p_1(\rrr)$ satisfies a Helmholtz equation with a complex-valued compressional wavenumber $\kc$ having the real part $k_0=\frac{\omega}{\cfl}$. We apply the effective boundary condition for $p_1(\rrr)$ recently derived by Bach and Bruus \cite{Bach2018}, which takes the viscous boundary layer into account at the domain boundary  $\pp\Omega$. We introduce the \textit{inward} normal derivative $\pp_\perp = -\nnn\cdot\nablabf$ and the effective \textit{inward} normal displacement $U_{1\perp}(\rrr)$ of the wall in terms of the actual displacement $\uuu_1(\rrr)$,
 \bsubal{p1_gov}
 \eqlab{p1_gov_helmholtz}
 &\lap p_1+\kc^2 p_1=0, \quad \kc
 =\Big(1+\ii\frac{\Gamfl}{2}\Big)k_0, \hspace{-2mm}  & \; &\hspace{-0mm}\rrr\in\Omega,
 \\
 \eqlab{p1_gov_bc}
 &\pp_\perp p_1+\frac{\ii}{\ks}\big(\kc^2 p_1+\pp_\perp^2 p_1\big)=\frac{ k_0^2U_{1\perp}(\rrr)}{\kapfl} ,  & &\hspace{-0mm}\rrr\in\pp \Omega,
 \\
 \eqlab{p1_gov_W}
 &U_{1\perp}(\rrr)= \frac{1}{1-\ii\Gamfl}\Big(-\nnn - \frac{\ii}{\ks}\nablabf\Big)\cdot\uuu_1, &
 &\hspace{-0mm}\rrr\in\partial\Omega.
 \esubal
Here, $\Gamfl$ is the weak bulk damping coefficient and $\ks$ is the viscous boundary layer wavenumber, given in terms of the viscous boundary-layer width $\delta$, the dynamic viscosity $\etafl$, the bulk viscosity $\etaB_\mr{fl}$, and the isentropic compressibility $\kapfl=\frac{1}{\rhofl}(\frac{\pp \rho}{\pp p})_S=\frac{1}{\rhofl\cfl^2}$ of the fluid,
\bal
 \eqlab{Gamfl_delta}
 \Gamfl=\bigg(\frac43+\frac{\etaBfl}{\etafl}\bigg)\etafl\kapfl\omega\ll1, \quad \ks=\frac{1+\ii}{\delta}, \quad \delta=\sqrt{\frac{2\etafl}{\rhofl\omega}}.
\eal

From the acoustic pressure $p_1$ we directly obtain the irrotational acoustic velocity $\vvv_1$ and the acoustic density $\rho_1$ in \eqref{pert_eqs} through the relations \cite{Bach2018},
 \bsubal{p_to_v_rho}
 \eqlab{p_to_v}
 \vvv_1&=-\frac{\ii(1-\ii\Gamfl)}{\omega\rhofl}\grad p_1,&  &\rot\vvv_1=\zerovec,\\
 \rho_1&=\rhofl \kapfl p_1.  &  &
 \esubal
Further, we define the following relevant time-averaged first-order products: the acoustic energy density $\Eac$, kinetic energy density $\Ekin$, potential energy density $\Epot$, energy flux density vector $\SSSac$, as well as the time-averaged acoustic angular momentum density $\Lac$ with respect to unperturbed fluid position $\rrr$,
 \bsubal{S}
 \Eac&=\Ekin+\Ekin, &\quad \Ekin&=\frac{1}{4}\rhofl |\vvv_1|^2, \\
 \SSSac&=\avr{p_1 \vvv_1},	& \quad  \Epot&=\frac{1}{4} \kapfl |p_1|^2,\\
 \eqlab{Lac}
 \Lac&=\avr{\ddd_1\times (\rhofl\vvv_1)}, &	 \quad  \ddd_1(\rrr)&=\frac{\ii}{\omega} \vvv_1(\rrr).
 \esubal
Here, $\ddd_1(\rrr)$ is the fluid displacement at position $\rrr$, and $\avr{A_1(\rrr,t)B_1(\rrr,t)}=\frac12 \Re [A_1(\rrr) B_1^*(\rrr)]$  is the usual time-average of products of time-harmonic fields $A_1$ and $B_1$, with the asterisk denoting complex conjugation.

\subsection{Acoustic streaming at fluid resonance}
Outside the narrow viscous boundary layers of width $\delta \lesssim 500\, \SInm$ (for water at MHz frequencies), the second-order acoustic streaming $\vvv_2$ is described as a Stokes flow driven by the acoustic body force $\fffac$ in the domain $\Omega$, see Eq.~(8) in Ref.~\cite{Karlsen2018} and Eq.~(52b) in Ref.~\cite{Bach2018}, and by the slip velocity $\vvv_2^\mr{slip}$ at the domain boundary $\pp \Omega$, see Eq.~(61) in Ref.~\cite{Bach2018},
 \bsublab{v2_gov}
 \bal
 \eqlab{v2_gov_cont}
 0 &= \div\vvv_2, &&\rrr\in \Omega,
 \\
 \eqlab{v2_gov_navier}
 \zerovec &= -\grad p_2
 + \etafl\lap\vvv_2
 +\fffac, &&\rrr\in \Omega,
 \\
 \eqlab{v2_gov_bc}
 \vvv_2&=\vvv_2^\mr{slip},  \qquad &&\rrr\in \pp \Omega.
\eal
Here, $\fffac$ in $\Omega$ and $\vvv_2^\mr{slip}$ at $\pp\Omega$, are given by
 \bal
 \eqlab{v2_gov_f_def}
 \fffac &=\frac{\Gamfl\omega}{\cfl^2}\SSSac,\\
 \eqlab{v2_gov_bc_def}
 \vvv_2^\mr{slip}&=\frac12\kapfl\SSSac-\frac{1}{2\rhofl\omega}\grad_\parallel\big[\Ekin-2\Epot\big]\nn\\
 &\quad +\frac{3}{2\omega}\avr{[\ii\pp_\perp v_{1\perp}]\vvv_{1\parallel}},
\eal
\esublab
where we have used the expression for the slip velocity $\vvv_2^\mr{slip}$ valid near a fluid resonance, in which case the magnitude $|\vvv_1|$ of the acoustic velocity $\vvv_1$ in the bulk is much larger than that of the wall velocity $\big|\omega \uuu_1\big| \ll |\vvv_1|$, see Ref.~\cite{Bach2018}. We call the streaming driven by the acoustic body force $\fffac$ in \eqref{v2_gov_f_def} for bulk-driven acoustic streaming, and the streaming driven by the slip velocity $\vvv_2^\mr{slip}$ in \eqref{v2_gov_bc_def} for boundary-driven acoustic streaming. Note that $\fffac$ increases with the frequency squared, as $\Gamfl\propto \omega$, whereas $\vvv_2^\mr{slip}$ is independent of the frequency, as $\grad_\parallel,\pp_\perp\sim k_0=\frac{1}{\cfl}\:\omega$.

One central quantity in the governing equations \eqnoref{v2_gov} for the acoustic streaming, is the acoustic energy flux density $\SSSac$, which enters both in the body force $\fffac$, \eqref{v2_gov_f_def}, and in the slip velocity $\vvv_2^\mr{slip}$, \eqref{v2_gov_bc_def}. This quantity is rotating if the acoustic fields are rotating, as seen by taking the curl of $\SSSac=\avr{p_1\vvv_1}$ and using  \eqref{p_to_v} with $\grad p_1\approx \ii\omega\rhofl\vvv_1$ and $\rot \vvv_1=\zerovec$,
\bsub\eqlab{rotS_intS}
 \bal \eqlab{rotS}
 \grad\times\SSSac=\omega^2 \Lac,
 \eal
where $\Lac$ is the time-averaged acoustic angular momentum density defined in \eqref{Lac}. Using Stokes theorem for any closed loop, shows that the energy flux density $\SSSac$, and therefore the body force $\fffac$, points in a direction which rotates around areas with high acoustic angular momentum density,
 \bal\eqlab{intS}
 \oint \fffac \cdot \dd \boldsymbol{l}
 = \frac{\Gamfl\omega}{\cfl^2} \oint \SSSac \cdot \dd \boldsymbol{l}
 = \frac{\Gamfl\omega^3}{\cfl^2}\int \Lac\cdot \nnn  \, \dd A.
\eal
\esub
Hence, bulk-driven streaming is a consequence of rotating acoustics with angular momentum density $\Lac$.

\section{Acoustic resonance modes in a box with viscous boundary layers}
\seclab{ModeTheory}
We consider the rectangular-box cavity of dimensions $\Lx\times\Ly\times\Lz,$ where $0\leq x\leq \Lx$, $0\leq y\leq \Ly$, and $0\leq z\leq \Lz$. Below, we give the solutions to \eqref{p1_gov} in this geometry for the resonance modes of the acoustic pressure with viscous boundary layers. We refer the reader to \appref{box_app} for the detailed derivations. We then evaluate the second-order quantities $\SSSac$, $\fffac$, and $\Lac$ which vanish for all single-mode resonances but not for double-mode resonances, where two single modes are excited simultaneously by the same frequency.

\subsection{Single-mode resonances}
\seclab{single_modes}

First, ignoring viscosity, the resonance solution for $p_1$ is proportional to the eigenfunctions $R^{lmn}(\rrr)$ of the Helmholtz equation \eqnoref{p1_gov_helmholtz}, which are the usual hard-wall standing resonance modes with integer number $l$, $m$, and $n$ half waves in the $x$, $y$, and $z$ direction, respectively,
 \bsub
 \eqlab{inviscid_modes}
 \bal \eqlab{def_Rlmn}
 p_1^{lmn}\propto R^{lmn}(\rrr)=\cos\big(\Kxnx x\big)\cos\big(\Kyny y\big)\cos\big(\Kznz z\big).
 \eal
Here, the wavenumbers $\Kxnx$, $\Kyny$, and $\Kznz$ are purely geometrical quantities that dictate the resonance wavenumbers $K_0^{lmn}$,
 \bal
 \Kxnx &= \frac{l \pi}{\Lx}, \quad
 \Kyny = \frac{m\pi}{\Ly}, \quad
 \Kznz = \frac{n\pi}{\Lz},\\
 K_0^{lmn} &= \sqrt{\big(\Kxnx\big)^2+\big(\Kyny\big)^2+\big(\Kznz\big)^2}.
 \eal
 \esub
In the presence of viscosity and hence viscous boundary layers, the resonant wavenumbers  $k_0^{lmn}$ and angular frequencies  $\omega^{lmn}$ are down-shifted slightly relative to the inviscid values due to the boundary-layer damping coefficients $\Gambl^{lmn}$,
 \bsubal{res_wave_numbers}
 \eqlab{k0lmn}
 k_0^{lmn} &= \Big(1-\frac12 \Gambl^{lmn}\Big)K_0^{lmn},
 \qquad \omega^{lmn} = k_0^{lmn} \cfl,\\
 \eqlab{Gambl}
 \Gambl^{lmn}&=
 \bigg(\frac{\Kxnx}{\KO^{lmn}}\bigg)^2\bigg(\frac{\delta}{\Lyny}+\frac{\delta}{\Lznz}\bigg)
 \nn \\
 & \hspace{1.5mm} + \bigg(\frac{\Kyny}{\KO^{lmn}}\bigg)^2\bigg(\frac{\delta}{\Lznz}+\frac{\delta}{\Lxnx}\bigg)\nn \\
  &\hspace{1.5mm} +\bigg(\frac{\Kznz}{\KO^{lmn}}\bigg)^2\bigg(\frac{\delta}{\Lxnx}+\frac{\delta}{\Lyny}\bigg),
 \esubal
where $\Lxnx=\int_0^{\Lx}\cos^2(\Kxnx x)\,\dd x =\frac12(1+\delta_{0l}) \Lx$ and similarly for $\Lyny$ and $\Lznz$. As an example, a pure $x$-mode will have $K_0^{l00}=\Kxnx$ and $\Ky^0=\Kz^0=0$ giving $\Gambl^{l00}=\frac{\delta}{\Ly}+\frac{\delta}{\Lz}$ corresponding to the boundary-layer damping from the four boundaries parallel to the $x$-axis. The total damping coefficient $\Gamma^{lmn}$ of the resonance mode $lmn$ includes both the boundary-layer-damping coefficients $\Gambl^{lmn}$ and the bulk damping coefficient $\Gamfl^{lmn}$, given in \eqref{Gamfl_delta} with $\omega=\omega^{lmn}$,
 \bal\eqlab{Gam_lmn}
 \Gamma^{lmn}=\Gambl^{lmn}+\Gamfl^{lmn}.
 \eal

For frequencies close to the resonance frequencies, we have $k_0 \approx k_0^{lmn}$ or $\omega \approx \omega^{lmn}$, and the resonance pressure mode $p^{lmn}_1$ satisfying \eqref{p1_gov} is the product of a complex-valued amplitude $\pa^{lmn}$, an internal frequency dependency $F^{lmn}(k_0)$, and the usual hard-wall spatial dependency $R^{lmn}(\rrr)$ from \eqref{def_Rlmn},
 \bsubalat{p_lmn_box}{3}
 \eqlab{p_lmn_box_PFR}
 p^{lmn}_1(\kO,\rrr)&= P^{lmn}_1 F^{lmn}(k_0)R^{lmn}(\rrr),\;  &&\text{ for } \kO\approx \kO^{lmn},\\
 \eqlab{p_lmn_box_Plmn}
 \pa^{lmn} &= \frac{\rhofl \cfl^2}{\Gamma^{lmn}}  \frac{\int_{\pp\Omega}
 U_{1\perp} R^{lmn}\ \mr{d} A}{\int_\Omega (R^{lmn})^2\, \dd V } &&= |\pa^{lmn} | \ee^{\ii\phiact^{lmn}},\\
 \eqlab{def_Flmn}
 F^{lmn}(k_0)&=\frac{\frac 12 k_0^{lmn}\Gamma^{lmn}}{(k_0-k_0^{lmn})+\ii \frac12 k_0^{lmn}\Gamma^{lmn}} && = |F^{lmn} | \ee^{\ii \phiF^{lmn}}.
 \esubalat
Here, we have also given $\pa^{lmn}$ and $F^{lmn}$ in their polar form and introduced the external actuation phase $\phiact^{lmn}$ as well as the internal frequency-dependent phase $\phiF^{lmn}$. The phase $\phiF^{lmn}$ is plotted in \figref{arg_and_phase_F} together with the modulus $|F^{lmn}|$. The total phase $\phitot^{lmn}$ of the mode $lmn$ is the sum of these phases,
 \beq{philmn}
 \phi_\mr{tot}^{lmn}=\phiact^{lmn}+\phiF^{lmn}.
 \eeq
 \begin{figure}[t]
\centering
\includegraphics[width=\columnwidth]{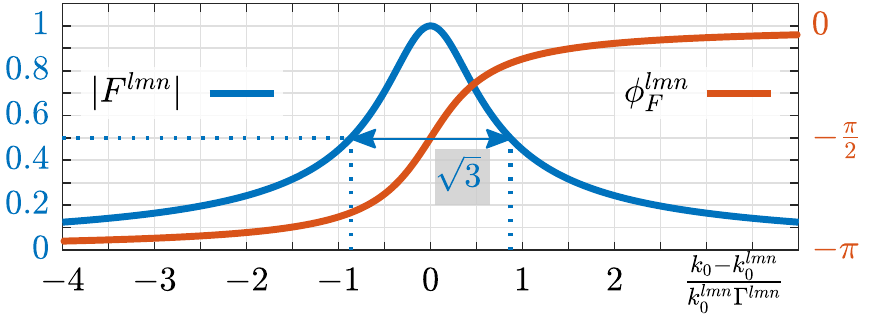}
\caption{\figlab{arg_and_phase_F} (Color online)
The modulus $|F^{lmn}|$ (left axis) and phase $\phiF^{lmn}$ (right axis) of the frequency-dependent factor $F^{lmn}$ in \eqref{def_Flmn}. The double arrow marks the full-width-at-half-maximum linewidth $\sqrt{3} k_0^{lmn}\Gamma^{lmn}$ of $|F^{lmn}|$.}
 \end{figure}

\subsection{Double-mode resonances}
\seclab{double_modes}

The bulk-driven acoustic  streaming is driven by the body force $\fffac$, defined in \eqref{v2_gov_f_def} and restated here solely in terms of the pressure $p_1$, by using $\vvv_1\approx \frac{-\ii}{\omega\rhofl}\grad p_1$ as stated in \eqref{p_to_v},
 \beq{fffacp1}
 \fffac=\Gamfl \kapfl \avr{[\ii p_1] \grad p_1}.
 \eeq
For any single-mode resonance $lmn$ of the form of \eqref{p_lmn_box_PFR}, this expression will vanish since $\ii p_1^{lmn}$ is exactly $\frac{\pi}{2}$ out of phase with
$\grad p_1^{lmn}$. This motivates us to consider double-mode resonances, which occur if two single modes $lmn$ and $\lmnp$ overlap, such that they can be excited simultaneously at the same frequency $f=\frac{1}{2\pi} k_0 \cfl$, with $k_0^{lmn} \approx k_0 \approx k_0^{\lmnp}$. In that case, the total pressure is the sum of two modes,
 \beq{p1lmnlmnp}
 p_1= p_{\lmnp}^{lmn} = p_1^{lmn}+p_1^\lmnp,
 \eeq
and the body force $\fffac$ in \eqref{fffacp1} becomes $\fff_{\lmnp}^{lmn}$,
 \bal
 \eqlab{fffac_lmnlmnp}
 \fff_{\lmnp}^{lmn}&=\Gamfl \kapfl \avr{[\ii (p_1^{lmn}+p_1^{\lmnp})] \grad (p_1^{lmn}+p_1^{\lmnp})}
 \nn \\
 &\hspace{-3mm}=\Gamfl \kapfl \big[\avr{[\ii p_1^{lmn}]\grad p_1^{\lmnp}}-\avr{p_1^{\lmnp}[\grad \ii p_1^{lmn}] }\big].
 \eal
Inserting the resonance~\eqnoref{p_lmn_box_PFR} with $\pa^{lmn}$ in polar form yields,
\bsub
 \eqlab{ALLlmnlmnp}
 \bal \eqlab{flmnlmnp}
 \fff_{\lmnp}^{lmn} &=\frac12 \Gamfl\kapfl \big|\pa^{lmn}\big|\big| \pa^{\lmnp}\big| \;
  \calF^{lmn}_\lmnp(k_0) \; \calKKK^{lmn}_\lmnp(\rrr), \\
  \eqlab{Flmnlmnp}
    \calF^{lmn}_\lmnp\,(k_0) &= \Re\Big\{\ii \Big[ F^{lmn}\ee^{\ii \phiact^{lmn}}\Big] \Big[F^{\lmnp}\ee^{\ii\phiact^{\lmnp}}\Big]^*\Big\},\\
    \eqlab{Klmnlmnp}
   \calKKK^{lmn}_\lmnp\,(\rrr) &=  R^{lmn}\grad R^{\lmnp}-R^{\lmnp}\grad R^{lmn}.
 \eal
\esub
Here, $\calF^{lmn}_\lmnp\,(k_0)$ gives the essential frequency dependency (as $\Gamfl$ is nearly constant over the width of the double mode), while $\calKKK^{lmn}_\lmnp\,(\rrr)$ with the dimension of a wave vector gives the spatial dependency and direction of the double-mode body force. Furthermore, we obtain the double-mode energy flux density $\SSS_{lmn}^\lmnp$ and acoustic angular momentum density $\calLLL^{lmn}_\lmnp$ directly from \eqref{flmnlmnp} by use of \eqsref{v2_gov_f_def}{rotS},
%
%\begin{widetext}
%\mbox{}\noindent
%\begin{figure}[h]
\begin{figure*}
\centering
\includegraphics[width=\textwidth]{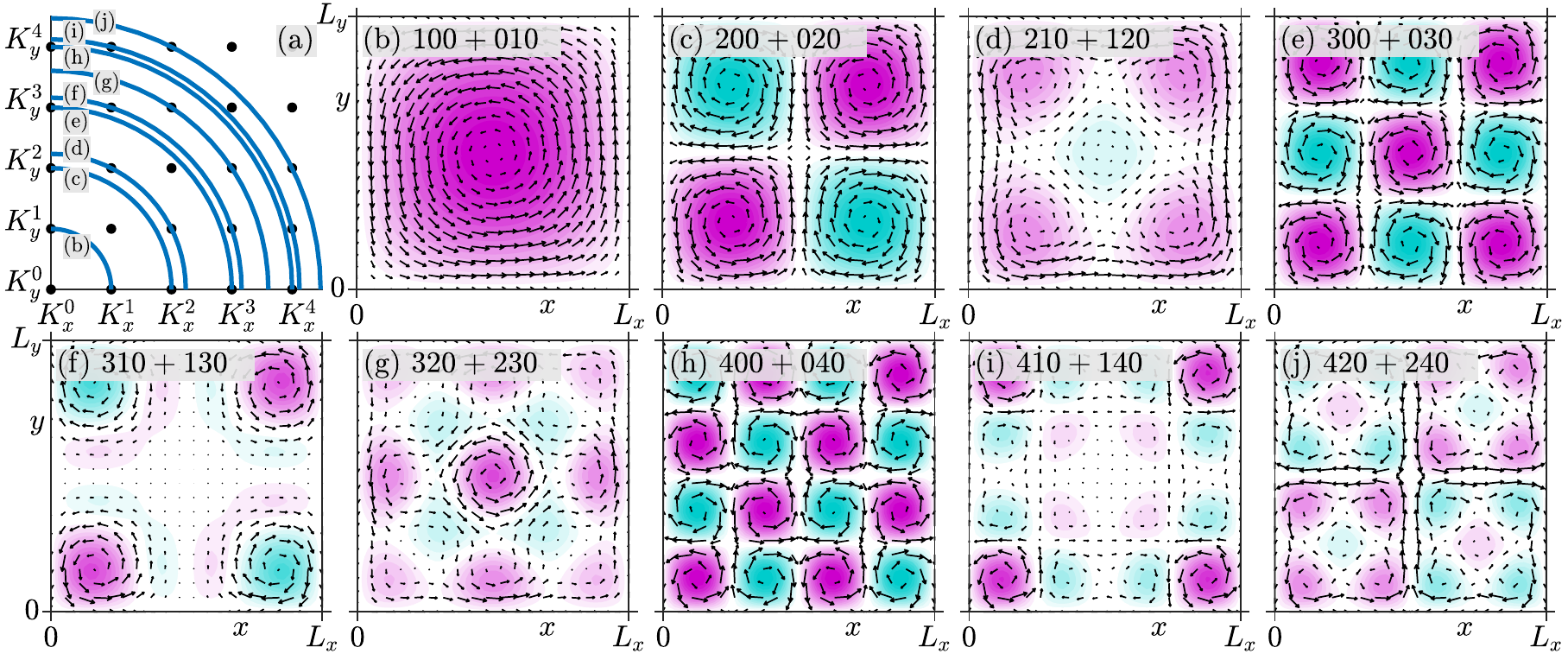}
\caption{\figlab{rotModes} (Color online)
Rotating acoustics in the $xy$-plane ($K^n_z = 0$) of a square cavity with $\Lx=\Ly$. (a) $K_x$-$K_y$-space diagram showing the allowed values of $K_x^l$ and $K_y^m$ (black dots). Each quarter circle (b)-(j) corresponds to the angular frequency $\cfl\sqrt{(K^l_x)^2+(K^m_y)^2}$, where the double mode $lm0+ml0$ is exited. (b)-(j) The corresponding acoustic rotation of  the double mode $lm0+ml0$, where the arrows represent the body force $\fff_{ml0}^{lm0}$ and the colors show the $z$-component of the spatial dependency $\rot \calKKK^{lm0}_{ml0}$ in \eqref{Llmnlmnp} of the acoustic angular momentum density from $-2k_0^2$ (light cyan) to $2k_0^2$ (dark magenta).}
\end{figure*}
%\end{figure}
%\end{widetext}
%
%
 \bsubal{Slmnlmnp_Llmnlmnp}
 \eqlab{Slmnlmnp}
 \SSS_{\lmnp}^{lmn} &= \frac12 \frac{ \abs{\pa^{lmn}} \abs{\pa^{\lmnp}}}{\rhofl\omega }  \calF^{lmn}_\lmnp(k_0) \calKKK^{lmn}_\lmnp(\rrr),\\
 \eqlab{Llmnlmnp}
 \calLLL^{lmn}_\lmnp &= \frac12 \frac{\abs{\pa^{lmn}} \abs{\pa^{\lmnp}} }{\rhofl\omega^3}  \calF^{lmn}_\lmnp(k_0) \rot\calKKK^{lmn}_\lmnp(\rrr).
 \esubal
Note that $\calF^{lmn}_\lmnp(k_0)$ is normalized to be between $-1$ and $1$, whereas $\calKKK^{lmn}_\lmnp(\rrr)$ and $\rot \calKKK^{lmn}_\lmnp(\rrr) = 2\grad R^{lmn}\times \grad R^{\lmnp}$ has the maximum amplitudes $k_0$ and $2 k_0^2$, respectively.

In \figref{rotModes} we show results for the nine lowest-frequency double-modes $lm0+ml0$ in a square cavity. We plot the body force $\fff^{lm0}_{ml0} \propto \calKKK^{lm0}_{ml0}$ and the $z$-component of the acoustic angular momentum $\calLLL^{lm0}_{ml0} \propto \rot \calKKK^{lm0}_{ml0}$. In the Supplemental Material \footnote{See Supplemental Material at [URL] extending the results of the following three figures: (1) \figref{rotModes} turned into one pdf-file with 55 sub-figures of the acoustic body force $\fffac$ of the double-modes $lm0+\lp\mp 0$ with $l,m,\lp,\mp = 0,1,2,3,4$ (the MATLAB source code is also provided); (2) \figref{p1_linePlots} turned into five animated gif files of the pressure $p_1$ in 3D and as line plots of the double modes $200+020$, $400+040$, and $600+060$ both for the wide $\frac12 \Lstar$-width and the narrow $\frac{1}{12}\Lstar$-width $G_1$ actuation; and (3) \figref{streaming246} re-calculated using the narrow $\frac{1}{12}\Lstar$-width $G_1$ actuation.}, we show $55$ examples of double-mode resonances $lm0+\lp\mp0$ with $l,m,\lp,\mp = 0,1,2,3,4$ in rectangular cavities with aspect ratios between $1$ and $4$, and we provide a Matlab code to compute these modes.

\section{Conditions for obtaining a strong acoustic body force}
\seclab{StrongBodyForce}
The magnitude of the body force $\fff_{\lmnp}^{lmn}$, the acoustic energy flux density $\SSS_{\lmnp}^{lmn}$, and the acoustic angular momentum density $\calLLL^{lmn}_\lmnp$ are all proportional to the frequency-dependent factor $\calF^{lmn}_\lmnp(k_0)$ in \eqref{Flmnlmnp}. To study this factor, we rewrite it to emphasize the total phase difference $\phi_\mr{tot}^{\lmnp}-\phi_\mr{tot}^{lmn}$ between the single mode phases given in \eqref{philmn},
\beq{calF_lmn_phase}
 \calF^{lmn}_\lmnp(k_0)=\big|F^{lmn}\big| \big|F^{\lmnp}\big|\sin(\phi_\mr{tot}^{\lmnp}-\phi_\mr{tot}^{lmn}).
\eeq
Consequently,  two conditions must be fulfilled to yield a strong body force: (\emph{i}) the overlap condition, \ie\ the two single-mode resonances must have a large overlap in frequency space to ensure a large value of the amplitude $\big|F^{lmn}\big| \big|F^{\lmnp}\big|$, and (\emph{ii}) the phase condition, \ie\ the difference $\phi_\mr{tot}^{\lmnp}-\phi_\mr{tot}^{lmn}$ in phases must be close to $\pm\frac{\pi}{2}$ to ensure a large value of the sine factor.

\subsection{The overlap condition: Aspect ratios}
\seclab{OverlapCond}

The condition that both $ |F^{lmn}|$ and $ |F^{\lmnp}|$ are large in \eqref{calF_lmn_phase} is satisfied if the modes overlap. This occurs if the resonance frequencies are nearly identical, $k_0^{lmn}\approx k_0^{\lmnp}$, or equivalently,
 \bal \eqlab{lm_lbmb_condition_3d}
 \Big(\frac{l\pi}{L_{x}}\Big)^2\hspace{-1mm}+\Big(\frac{m\pi}{L_{y}}\Big)^2\hspace{-1mm}+\Big(\frac{n\pi}{L_{z}}\Big)^2 \approx \Big(\frac{\lp\pi}{L_{x}}  \Big)^2\hspace{-1mm}+\Big(\frac{\mp\pi}{L_{y}}\Big)^2\hspace{-1mm}+\Big(\frac{\np\pi}{L_{z}}\Big)^2,
 \eal
where we have used that $\Gambl^{lmn}\ll 1$ in \eqref{k0lmn} so $k_0^{lmn}\approx K_0^{lmn}$. In the general case, \eqref{lm_lbmb_condition_3d} gives a myriad of possible mode combinations. If we restrict ourselves to horizontal modes, where $n=\np=0$, we can solve for the aspect ratio $A=\frac{\Lx}{\Ly}$ in \eqref{lm_lbmb_condition_3d},
 \bal\eqlab{Asqr_condition}
 A=\frac{\Lx}{\Ly} \approx \sqrt{\frac{\lp^2-l^2}{m^2-\mp^2}}.
 \eal
Listing the integers $l,m,\lp,\mp$ for which $A$ is real, we find that the square cavity with $A\approx 1$ is the richest case having the highest number of allowed double-mode excitations. In \figref{rotModes}, we show the first 9 horizontal double-mode excitations in a square cavity with $A=1$ and $n=\np=0$. In this case, the condition in \eqref{lm_lbmb_condition_3d} corresponds to two points $\big(\frac{l\pi}{\Lx},\frac{m\pi}{\Ly}\big)$ and $\big( \frac{\lp\pi}{\Lx},\frac{\mp \pi}{\Ly} \big)$ both lying on the same quarter circle of radius $K_0$, as illustrated by the black dots in \figref{rotModes}(a).

\subsection{The phase condition: Rotating versus non-rotating actuation}
\seclab{PhaseCond}

The straightforward way to obtain rotating acoustics is to actuate a given system with two transducers running with a phase difference of  $\phiact^{\lmnp} - \phiact^{lmn} = \pm\frac{\pi}{2}$, a technique that has been used in many experiments to generate acoustic radiation torques \cite{Busse1981, Tran2012, Dual2012d, Lee1989, Courtney2013, Mao2017a}. Remarkably, even in the standard case of microfluidic ultrasound experiments driven by a single piezoelectric transducer with a non-rotating actuation, $\phiact^{\lmnp} - \phiact^{lmn}=0$, a rotating acoustic field may nevertheless be generated because the total phase difference $\phitot^{\lmnp} - \phitot^{lmn} = \phiF^{\lmnp} - \phiF^{lmn}$ may be close to $\pm\frac{\pi}{2}$ due to the internal phases $\phiF^{lmn}$ of the frequency-dependent factors $F^{lmn}(k_0)$, see \eqref{def_Flmn} and \figref{arg_and_phase_F}.

\section{Non-rotating actuation of rotating double-mode resonances in a nearly-square cavity}
\seclab{NonRotActuation}

We now show how a non-rotating actuation can lead to a rotating double-mode resonance in a nearly-square cavity characterized by the  average side length  $\Lstar=\frac12 (\Lx+\Ly)$ and the small symmetry breaking $d=\Lx-\Ly$,
 \beq{L_perturb}
 \Lx=\Lstar+\frac12 d, \quad \Ly=L^{\star}-\frac12 d,
 \quad \text{with } |d| \ll \Lstar.
 \eeq
The magnitude of the resulting acoustic body force $\fffac$ depends strongly on the symmetry breaking $d$, and for a given double mode we determine how much the aspect ratio $A = L_x/L_y \approx 1 + d/\Lstar$ should deviate from unity to maximize $\fffac$. Specifically, we consider the symmetric double-mode excitations $lm0+ml0$ in the horizontal $xy$ plane of the type shown in \figref{rotModes}. For such double-modes, the analysis simplifies significantly, not only because $z$ drops out, but also because the damping coefficients $\Gamma^{lm0}$ and $\Gamma^{ml0}$ from \eqsref{Gambl}{Gam_lmn} are almost identical. In short, we consider the situation,
\bsubal{nearly_square_conditions}
 &\text{Nearly square cavity:}  & &\Lx \approx \Lstar \approx \Ly,\\
 &\text{Horizontal $xy$-modes:} & &n=\np=0 ,\\
 &\text{Symmetric double modes:}& &\lmp=ml ,\\
 &\text{Non-rotating actuation:}& & \phiact^{lm0}=\phiact^{ml0} ,\\
 &\text{Similar damping coefficients:} & & \Gamma^{lm0}\approx \Gamma^{ml0}.
\esubal
To analyse the strength of the acoustic body force, we examine the factor $\calF_{lm0}^{ml0}(k_0)$ in \eqref{Flmnlmnp} that describes the frequency-dependency of the acoustic body force. For the double mode $lm0+ml0$, we introduce the short two-subscript notation $\calF_{lm}(k_0)$ for this quantity,
 \bal \eqlab{non-rotation_calF}
 \calF_{lm}(k_0)&=\calF_{ml0}^{lm0}(k_0)=\Re\Big\{\ii \Big[ F^{lm0}\Big] \Big[F^{ml0}\Big]^*\Big\}\nn \\
 &=|F^{lm0}| |F^{ml0}| \sin\big(\phi_F^{ml0}-\phi_F^{lm0}\big),
 \eal
where the functions $F^{lm0}(k_0)$ and $F^{ml0}(k_0)$ behave as shown in \figref{arg_and_phase_F}. Note that if the modes $lm0$ and $ml0$ have equal resonance frequencies, they have full mode overlap, and the product $|F^{lm0}| |F^{ml0}|$ is maximized, while $\sin\big(\phi_F^{ml0}-\phi_F^{lm0}\big)$ vanishes, and there is no body force. To maximize $\calF_{lm}(k_0)$, the two modes must therefore be separated sufficiently to ensure that the phase difference $\phi_F^{ml0}-\phi_F^{lm0}$ deviates from zero, but not too much, as a decreased mode overlap reduces the modulus product $|F^{lm0}| |F^{ml0}|$. Therefore, we cannot in this case perfectly fulfil both the overlap condition, \secref{OverlapCond}, and the phase condition, \secref{PhaseCond}, so the strongest body force is found at a certain mode separation $\Delta_{lm}=k_0^{lm0}-k_0^{ml0}$ between the resonance wavenumbers, which gives only partial mode overlap and a phase difference that deviates from the ideal $\pm\frac{\pi}{2}$.

The mode separation $\Delta_{lm}$ is induced by the symmetry breaking $d$, and similar to \eqref{L_perturb}, we introduce the center wavenumber $k^\star_{lm} =\frac12 (k_0^{lm0}+k_0^{ml0})$ corresponding to the center frequency between the two modes and write,
 \bsub
 \bal
 \eqlab{k_perturb}
 k_0^{lm0} &=k_{lm}^{\star}+\frac12 \Delta_{lm}, \qquad k_0^{ml0}=k_{lm}^{\star}-\frac12 \Delta_{lm},\\
 \eqlab{kStarlm}
 k^\star_{lm} &
 \approx \Big(1-\frac12 \Gambl^{lm0}\Big) \sqrt{l^2+m^2}\:\frac{\pi}{\Lstar}.
 \eal
  \esub
The relation between the mode separation $\Delta_{lm}$, the symmetry breaking $d$, and the aspect ratio $A=\Lx/\Ly$, is found by inserting \eqref{L_perturb} into $k_0^{lm0} = \sqrt{(\frac{l\pi}{\Lx})^2+ (\frac{m\pi}{\Ly})^2}+\ord{\Gamma^{lm0}}$,  expanding in the small ratio $\frac{d}{\Lstar}$, and comparing with \eqref{k_perturb},
 \beq{d_to_k_perturb}
 \Delta_{lm} \approx \frac{m^2-l^2}{\sqrt{m^2+l^2}}\frac{\pi}{\Lstar} \frac{d}{\Lstar}\approx \frac{m^2-l^2}{\sqrt{m^2+l^2}}\frac{\pi}{\Lstar} (A-1).
 \eeq
\begin{table}[b!]
\centering
\caption{\tablab{Gamma_list} The damping coefficients $\Gamma$, the full-width-half-maximum (FWHM) line widths, and the mode separation $\Delta$.}
\begin{ruledtabular}
\begin{tabular}{lcc}
Parameter & Symbol & Reference\\
\hline
Bulk damping & $\Gamfl$   &  \eqref{Gamfl_delta} \rule{0ex}{3.0ex}\\
Boundary-layer damping & $\Gambl^{lmn}$   &  \eqref{Gambl}\\
Total damping & $\Gamma^{lmn}$   &  \eqref{Gam_lmn}\\
FWHM of $\big|F^{lmn}\big|$ & $\sqrt{3}k_0^{lmn}\Gamma^{lmn}$ & \figref{arg_and_phase_F}\\
FWHM of $\Eac^{lmn}\propto \big|F^{lmn}\big|^2$ & $k_0^{lmn}\Gamma^{lmn}$ &
Eqs.~\eqnoref{S},\eqnoref{p_lmn_box}\\
Optimal mode separation $\Delta_{lm}^\mr{opt}$ & $\frac{1}{\sqrt{3}}\:k^\star_{lm}\Gamma^\star_{lm}$ & \eqref{Delta_opt}
\end{tabular}
\end{ruledtabular}
\end{table}
Since the damping coefficients are nearly identical $\Gamma^{lm0}\approx \Gamma_{lm}^\star \approx \Gamma^{ml0}$, with $\Gamma_{lm}^\star=\frac12 (\Gamma^{lm0}+\Gamma^{ml0})$ being the average value, it is convenient to scale all wavenumber quantities by their common line width $\Gamma^\star_{lm} k_{lm}^\star$, see \tabref{Gamma_list},
 \beq{ktilde_def}
 \tilde{k}_0=\frac{k_0}{k_{lm}^\star\Gamma^\star_{lm}}, \quad \kti_{lm}^ \star=\frac{k_{lm}^\star}{k_{lm}^\star\Gamma^\star_{lm}}, \quad \tilde{\Delta}_{lm} =\frac{\Delta_{lm}}{k_{lm}^\star\Gamma^\star_{lm}}.
 \eeq
In terms of these quantities, we write the frequency-dependencies  $F^{lm0}(k_0)$ and $F^{ml0}(k_0)$ from \eqref{def_Flmn} as,
 \bsubal{Flmn_normalized}
 F^{lm0}(k_0)&=\frac{1}{2(\kti_0-\kti_{lm}^\star)-\Deltati_{lm}+\ii},\\
 F^{ml0}(k_0)&=\frac{1}{2(\kti_0-\kti_{lm}^\star)+\Deltati_{lm}+\ii}.
 \esubal
\begin{figure}[b!]
\centering
\includegraphics[width=\columnwidth]{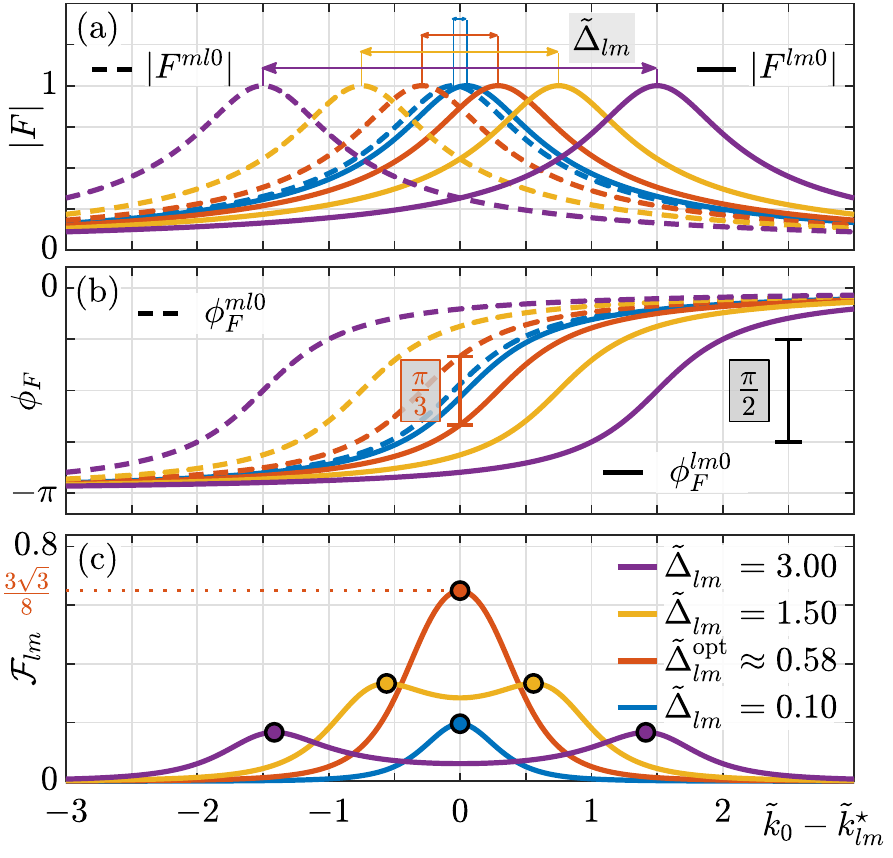}
\caption{\figlab{NonRotActuation}  (Color online)
The mechanism behind rotating acoustics for non-rotating actuations. (a) The modulus $|F|$ and (b) the phase $\phi_F$ of the mode $lm0$ (left, dashed) and $ml0$ (right, solid) given in \eqref{Flmn_normalized} for different mode separations $\tilde{\Delta}_{lm}$. At the optimal mode separation $\Deltati_{lm}^\mr{opt}=\frac{1}{\sqrt{3}}\approx 0.58$, the phase difference at the center frequency is $\frac{\pi}{3}$. (c) The frequency-dependent factor $\calF_{lm}(k_0)$, from \eqref{non-rotation_calF}, which describes the strength of the acoustic body force $\fffac$, when the modes of same color in (a) and (b) are combined. The black-edged circles in (c) correspond to those of the same color in \figref{rot_modes_freq_dep_d}. The legends in panel (c) apply also for (a) and (b).}
 \end{figure}
In \figref{NonRotActuation}(a) and (b) we show as in \figref{arg_and_phase_F} the modulus and phase, respectively, of $F^{lm0}$ and $F^{ml0}$ for four different mode separations $\Deltati_{lm} = 0.10, 0.58, 1.50, 3.00$. In \figref{NonRotActuation}(c) we show for the same values of $\Deltati_{lm}$ the frequency dependency $\calF_{lm}(k_0)$ from \eqref{non-rotation_calF} of the acoustic body force. We note that there exist an optimal mode separation $\Deltati_{lm}^{\mr{opt}}$ that yields the largest possible value  $\calF_{lm}^\mr{opt}$ of $\calF_{lm}$ found at the optimal wave number $k_0^\mr{opt}$. In \eqref{Deltati_opt_appendix} of \appref{Deltaopt_app} we show that
 \bsub
 \eqlab{geom_opt}
 \balat{2}  \eqlab{k0_opt}
  k_0^\mr{opt}  &= k^\star_{lm}, &&\\
 \eqlab{Delta_opt}
 \Deltati_{lm}^{\mr{opt}} &= \pm\frac{1}{\sqrt{3}} && = \pm 0.58,\\
 \eqlab{Flm_opt}
 \calF_{lm}^\mr{opt} &= \pm \frac{3\sqrt{3}}{8} && = \pm 0.65.
 \ealat
This means that the optimal frequency is the center frequency between the modes, and the optimal mode separation is $\frac{1}{\sqrt{3}}$ of the full-width-half-maximum. Inserting the optimal conditions \eqnoref{k0_opt} and \eqnoref{Delta_opt} into \eqref{Flmn_normalized} reveals the phase difference between the modes at the optimal conditions,
\bal
 \big(\phi_F^{ml0}-\phi_F^{lm0}\big)^\mr{opt} &= \pm\frac13\:\pi.
\eal
By using \eqref{d_to_k_perturb}, we translate the optimal mode separation $\Deltati_{lm}^\mathrm{opt}$ into the optimal size of the symmetry breaking $d_{lm}^\mathrm{opt}=(\Lx-\Ly)_{lm}^\mathrm{opt}$ and the corresponding aspect ratio $A_{lm}^\mathrm{opt}=(\frac{\Lx}{\Ly})_{lm}^\mathrm{opt}$, where the acoustic body force $\fffac$ of the double mode $lm0+ml0$ is maximized,
 \bal
 \eqlab{d_opt}
 d^\mr{opt}_{lm}&= \Deltati_{lm}^{\mr{opt}}\frac{l^2+m^2}{m^2-l^2}
 \Gamma^\star_{lm}\Lstar,\\
 \eqlab{A_opt}
 A^\mr{opt}_{lm}&=1+ \frac{1}{\Lstar }d^\mr{opt}_{lm}.
 \eal
 \esub
For all the double modes shown in \figref{rotModes}, we have $l>m$, so for a non-rotating actuation, the rotation direction of the body force will therefore be as shown in the case of $d<0$ (corresponding to $\Lx<\Ly$ and $A<1$), and opposite for $d>0$. Note from \eqref{d_opt} that this asymmetry-induced rotating acoustics is extremely sensitive to  $d$. In practise, it may be difficult to control this mechanism, as it requires a fabrication accuracy much smaller than $d/\Lstar \sim \Gamma_{lm}^\star$, which in microfluidic devices typically is smaller than 1\% \cite{Hahn2015}.

\section{3D simulation of acoustic streaming in a nearly-square cavity}
\seclab{3DSimulation}

In this section we validate by direct numerical simulation the theory for rotating acoustics induced by a non-rotating actuation in a weakly symmetry-broken geometry. We simulate the acoustic pressure $p_1$ and the acoustic streaming $\vvv_2$ in a shallow, nearly-square cavity in three dimensions (3D).  We choose a cavity similar to that investigated experimentally by Hags\"ater \etal~\cite{Hagsater2007}, and simulated in 3D using the hard-wall approximation by Lei \etal~\cite{Lei2014}, and including the actuator and elastic walls by Skov \etal~\cite{Skov2019}.  The cavity has the mean side length $\Lstar=2000\, \SIum $, $\Lx=\Lstar+\frac12 d$, $\Ly = \Lstar-\frac12 d$, and height $\Lz=200\, \SIum$. We use \eqref{geom_opt} to optimize the symmetry breaking $d$ and actuation frequency $f=\frac{1}{2\pi}k_0 \cfl$ to obtain the strongest possible rotating acoustic streaming. The coordinate system is the same as in \secref{ModeTheory}, with $0\leq x\leq \Lx$, $0\leq y\leq \Ly$, and $0\leq z\leq \Lz$. The boundary is taken to be stationary everywhere, except at the bottom boundary $z=0$, where we apply the non-rotating Gaussian wall displacement  in the normal direction $z$, with maximum amplitude $u_{1z}^0$ in the cavity center and minimum amplitude $\ee^{-2}u_{1z}^0$ in the corners,
 \bal \eqlab{G1_actuation}
 \uuu_1 &= \left\{ \begin{array}{cl}
 u_{1z}^0 G_1(x,y)\:\eee_z, & \text{ for }\; z = 0,\\
 \zerovec, & \text{ otherwise}, \end{array} \right. \\[2mm]
 G_1(x,y) &=
 \exp\Bigg[-\frac{\big(x-\frac{L_x}{2}\big)^2}{\big(\frac{L_x}{2}\big)^2}
 -\frac{\big(y-\frac{L_y}{2}\big)^2}{\big(\frac{L_y}{2}\big)^2}\Bigg].  \nn
 \eal
In Section~\secrefnoSec{ResultPressure}--\secrefnoSec{ResultBulkAna} the fluid is water, and in \secref{ResultBulkVisc} it is pyridine. The material parameters for these fluids are listed in \tabref{fluid_params}.
\begin{table}[t!]
\centering
\caption{\tablab{fluid_params} Material parameters for water \citep{Muller2014} and pyridine (C$_5$H$_5$N) \citep{Dukhin2009} at 25 C$^\circ$ used in the numerical simulations.}
\begin{ruledtabular}
\begin{tabular}{lcrrl}
Parameter & Symbol & Water & Pyridine & {Unit}  \\ \hline
Mass density     &   $\rhofl$ & 997.05&982 & kg~m$^{-3}$ \rule{0ex}{3.0ex} \\
Compressibility   & $\kapfl$ & 448  & 507& TPa$^{-1}$  \\
Speed of sound  & $\cfl$ &1496.7 & 1417& m~s$^{-1}$ \\
Dynamic viscosity & $\etafl$&  0.890 & 0.879& mPa\,s \\
Bulk viscosity & $\etaBfl$ &2.485 &  62.4 & mPa\,s
\end{tabular}
\end{ruledtabular}
\end{table}

In the following simulations, we excite the double-mode resonance $l00$+$0l0$ with $l=2,4,$ and $6$. Using our analytical expressions \eqrefnoEq{ALLlmnlmnp}, we have in \figref{rotModes} shown the spatial dependency of  the acoustic  body force $\fffac = \fff^{l00}_{0l0}$ for these particular double modes. As seen in \eqref{v2_gov}, this body force is the source of the bulk-driven acoustic streaming, so we predict that the numerical simulation of the bulk-driven streaming will result in a 2-by-2 pattern as in \figref{rotModes}(c) for $l=2$, a 4-by-4 pattern as in \figref{rotModes}(h) for $l=4$, and similarly for $l=6$ (not shown in \figref{rotModes}).

\subsection{Choosing the parameters}

We derived in \secref{ModeTheory} analytical expressions for the acoustic resonance modes including the viscous boundary layers, and we derived in \secref{NonRotActuation} the conditions on the geometry of a nearly-square cavity for which the body force $\fffac$ and the acoustic rotation are maximized for a non-rotating actuation. Remarkably, these analytic results allow an analytical determination of the parameter values that lead to the strongest rotating acoustic streaming in the $xy$-plane.

For the double mode $l00$+$0l0$, the small damping coefficients $\Gamfl^{l00}\approx \Gamfl^{0l0}$ in \eqref{Gamfl_delta} and $\Gambl^{l00}\approx \Gambl^{0l0}$ in \eqref{Gambl} are computed by approximating the angular frequency by the constant value $\omega = \cfl k_{l0}^\star\approx \cfl \frac{l\pi}{\Lstar}$, see \eqref{kStarlm},
 \bsubal{Gamma_l00}
 \eqlab{Gamfl_l00}
 \Gamfl^{l00} &\approx \Big(\frac{4}{3}+\frac{\etaBfl}{\etafl}\Big)\frac{\etafl}{\rhofl\cfl} \frac{l\pi}{\Lstar}\approx \Gamfl^{0l0},\\
 \eqlab{Gambl_l00}
 \Gambl^{l00} &\approx \frac{\delta}{\Lstar}+\frac{\delta}{L_z}\approx \Gambl^{0l0}, \qquad \delta\approx\sqrt{\frac{2\etafl \Lstar}{l\pi \cfl\rhofl}}.
 \esubal
Combining Eqs.~\eqrefnoEq{k0lmn}, \eqrefnoEq{k0_opt}, and \eqrefnoEq{kStarlm}, we obtain an expression for the boundary-layer shifted, optimal double-mode actuation frequency, which is the center frequency of the two single modes,
 \beq{f_l00}
 f_{l0}^\star \approx \frac{l}{2} \Big(1-\frac12 \Gambl^{l00}\Big)  \frac{\cfl}{\Lstar}.
 \eeq
\begin{table}[t!]
\caption{Parameters used in the numerical simulations of the double modes $l00+0l0$. The geometry parameters are optimized for maximum acoustic rotation in water, and the resulting dependent parameters (damping factors and mode amplitudes) are listed for both water and pyridine.}
\tablab{params246}
\begin{ruledtabular}
\begin{tabular}{l c c r r r}
Parameter						& 	See Eq.			& Unit									 &$l=2$ 		 & $l=4$ 		 & $l=6$\\
\hline
\multicolumn{6}{l}{\textit{Geometry (optimized for water) \rule{0ex}{3.0ex}}}\\
$d_{l0}^\mr{opt}$	&	\eqnoref{d_opt_l00} &   $\SIum	$ 						& \hspace{-1mm} $-3.922$ 	& \hspace{-1mm}$-2.785$	& \hspace{-1mm} $-2.286$ \\
$A_{l0}^\mr{opt}$	&	\eqnoref{A_opt_l00} &   1 						& 0.9980 	& 0.9986	& 0.9989 \\
$\Lstar$	&	 &   $\SIum	$ 						& \phantom{.}2000 	& \phantom{.}2000	 & \phantom{.}2000 \\
$L_z$	&	 &   $\SIum	$ 						& \phantom{2.}200 	& \phantom{2.}200	& \phantom{2.}200 \\
\multicolumn{6}{l}{\textit{Dependent parameters (water) \rule{0ex}{3.0ex}}}\\
$u^0_{1z}$			& \eqnoref{uwall_l00}	&$ \SInm$ 		& 1.450	& 7.474	& \hspace{-2.7mm} 14.027 \\
$f_{l0}^\star$ 		&\eqnoref{f_l00}	&$ \SIMHz$ 	 & 0.747 	& 1.494 	& 2.243 \\
$\Gamfl^{l00}\approx\Gamfl^{0l0}$ 	 &	\eqnoref{Gamfl_l00}	& $10^{-3}$				& 0.008 	 & 0.015	& 0.023 \\
$\Gambl^{l00}\approx\Gambl^{0l0}$  &	\eqnoref{Gambl_l00} & $10^{-3}$					& 3.389 	 & 2.396	 & 1.957 \\
\multicolumn{6}{l}{\textit{Dependent parameters (pyridine)\rule{0ex}{3.0ex}}}\\
$u^0_{1z}$			& \eqnoref{uwall_l00}	&$ \SInm$ 		& 1.757	& 9.665	& \hspace{-2.7mm} 19.617 \\
$f_{l0}^\star$ 		&\eqnoref{f_l00}	&$ \SIMHz$ 	 & 0.707 	& 1.415 	& 2.123 \\
$\Gamfl^{l00}\approx\Gamfl^{0l0}$ 	 &	\eqnoref{Gamfl_l00}	& $10^{-3}$				& 0.144 	 & 0.287	& 0.431 \\
$\Gambl^{l00}\approx\Gambl^{0l0}$  &	\eqnoref{Gambl_l00} & $10^{-3}$					& 3.488 	 & 2.467	 & 2.014 \\
\end{tabular}
\end{ruledtabular}
\end{table}
The optimal symmetry breaking $d_{l0}^\mr{opt}$ and the corresponding aspect ratio  $A_{l0}^\mr{opt}$ then follow from \eqsref{d_opt}{A_opt},
 \bsubal{dA_opt_l00}
 \eqlab{d_opt_l00}
 d_{l0}^\mr{opt} &\approx \frac{-1}{\sqrt{3}} \Big(\Gambl^{l00}+\Gamfl^{l00}\Big) \Lstar,\\
 \eqlab{A_opt_l00}
 A_{l0}^\mr{opt} &\approx 1-\frac{1}{\sqrt{3}} \Big(\Gambl^{l00}+\Gamfl^{l00}\Big).
 \esubal
Finally, we set the pressure amplitudes $\pa^{l00}=\pa^{0l0}=  1~\SIMPa$ in \eqref{p_lmn_box_Plmn}, and compute the corresponding amplitude $u^0_{1z}$ of the actuation displacement in \eqref{G1_actuation} as,
 \beq{uwall_l00}
 u_{1z}^0=\dfrac{\Lx\Ly\Lz(\Gamfl^{l00}+\Gambl^{l00}) \kapfl \pa^{l00}}{2\int_0^{\Lx}\int_0^{\Ly} G_1(x,y)\cos (\frac{l\pi}{L^{\star}}x)\,\dd x\dd y}.
 \eeq
For water and $l=2$, 4, and 6, the values of the expressions in \eqref{Gamma_l00}-\eqnoref{uwall_l00} are listed in \tabref{params246}. For pyridine, we adjust the actuation amplitude and actuation frequency to maintain the resonance amplitudes of $\pa^{l00}=\pa^{0l0}=  1~\SIMPa$. Note that for pyridine, we use the same optimal symmetry breaking $d_{l0}^\mr{opt}$ as used for water. By combining \eqssref{non-rotation_calF}{Flmn_normalized}{d_opt} with the material parameters in \tabref{fluid_params}, this leads to the values $\calF_{l0}=0.65$, 0.64, and 0.63 for $l=2$, 4, and 6, respectively, which are nearly the same as the optimal value $\calF_{l0}^\mr{opt}=0.65$ given in \eqref{Flm_opt}.

\subsection{Implementation in COMSOL}

\vspace*{-4mm}

As in our previous work \cite{Muller2012, Ley2016, Skov2019}, the simulation is performed in two steps using the \textit{Weak Form PDE} module in the finite-element-method software COMSOL Multiphysics \cite{COMSOL54}. In the first step, we solve for the acoustic pressure $p_1$ by implementing \eqref{p1_gov} in weak form, and in the second step we solve \eqref{v2_gov} for the steady streaming $\vvv_2$ and pressure $p_2$. The zero-level of $p_2$ is fixed by using the \textit{Global Constraint} $\int_\Omega p_2 \, \dd V=0$. For numerical efficiency, we exploit the symmetry at the vertical planes $x=\frac12 \Lx$ and $y=\frac12 \Ly$, where we apply the symmetry conditions $\nnn\cdot\grad p_1=0$, $\nnn\cdot\vvv_2=0$, and $\nnn\cdot\grad \vvv_{2\parallel}=\zerovec$. The computational domain is thus reduced to the quadrant $0\leq x \leq \frac12 L_x$ and $0\leq y \leq \frac12 L_y$. We use quartic-, cubic-, and quadratic-order Lagrangian shape functions for $p_1$, $\vvv_2$, and $p_2$, respectively, and with a tetrahedral finite-element mesh of mesh size $\frac{1}{6} \Lz$, this leads to $1.3\times 10^6$ degrees of freedom and a relative accuracy better than 1~\%. The simulations were performed on a workstation with a 3.5-GHz Intel Xeon CPU E5-1650 v2 dual-core processor, and with a memory of 128~GB RAM.

\subsection{Simulation results}
\seclab{SimulationResults}

\subsubsection{The acoustic pressure}
\seclab{ResultPressure}

In \figref{p1_linePlots}, we show the simulation results $p_1^\mr{num}$ of the acoustic pressure $p_1$ in water for $l=6$ in 3D and lineplots for $l=2,4$, and $6$. These three modes are examples of rotating double modes actuated by the wide, single, non-rotating Gaussian actuation $G_1$, \eqsref{G1_actuation}{uwall_l00}, at the bottom boundary $z=0$. In the Supplemental Material~\cite{Note1} we show the rotating dynamics of the pressure contours similar to \figref{p1_linePlots}(a) for $l=2,4,$ and $6$ as gif animations. In \figref{p1_linePlots} we also show the the analytical result $p_1^\mr{ana} = p_{0l0}^{l00} = p_1^{l00}+p_1^{0l0}$, which is obtained from \eqsref{p_lmn_box}{p1lmnlmnp} with the optimal mode separation $k_0^{l00}-k_0^{0l0}=\frac{1}{\sqrt{3}} k_{l0}^\star \Gamma_{l0}^\star$, \eqref{Delta_opt}, and mode amplitudes $\pa^{l00}=\pa^{0l0}=1\, \SIMPa$,
 \begin{figure}[t]
\centering
\includegraphics[width=\columnwidth]{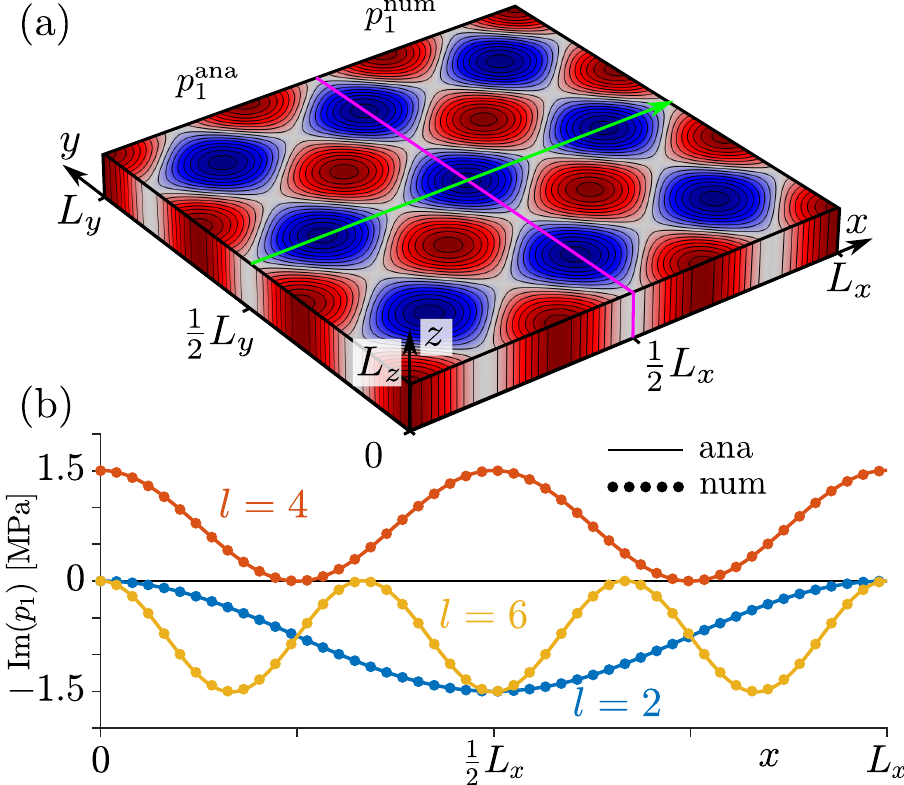}
\caption{\figlab{p1_linePlots}  (Color online)
The acoustic pressure $p_1$ for water in the nearly-square cavity with parameters given in \tabsref{fluid_params}{params246}, where the double mode $l00+0l0$ is excited by the non-rotating actuation $G_1$, \eqsref{G1_actuation}{uwall_l00}. (a) Pressure contour plot from $-1.5$~MPa (dark blue) to $1.5$~MPa (light red) for $l=6$ of the analytic solution $p_1^\mathrm{ana}$, \eqref{p_l00s_0l0_sim} (for $x<\frac12 \Lx$), and the numerical solution $p_1^{\mr{num}}$ (for $x>\frac12 \Lx$). (b) Line plots of the pressure $p_1^\mathrm{ana}$ (solid) and $p_1^{\mr{num}}$ (dots) along the green line in (a) at $y=\frac12 \Ly$ and $z=\Lz$ for the modes $l=2$ (blue), $l=4$ (orange), and $l=6$ (yellow). Animations of the rotating acoustic fields are shown in the Supplemental Material~\cite{Note1}.}
 \end{figure}
 \beq{p_l00s_0l0_sim}
 p_1^\mr{ana} = p_{0l0}^{l00}
 = 1\ \SIMPa\times \Bigg[ \frac{\cos\Big(\frac{l\pi x}{\Lstar}\Big)}{-\frac{1}{\sqrt{3}}+\ii} +\frac{\cos\Big(\frac{l\pi y}{\Lstar}\Big)}{\frac{1}{\sqrt{3}}+\ii}  \Bigg].
 \eeq
In \figref{p1_linePlots}(b), we compare $\im\big(p_1^\mr{num}\big)$ and $\im\big(p_1^\mr{ana}\big)$ and find a quantitative agreement better than 1~\%. The reason for choosing the imaginary part is to single out the resonant part of the field, which is phase shifted by the factor $\exp\big(\ii\frac{\pi}{2}\big) = \ii$ relative to the actuation displacement $\uuu_1$ at the boundary, \eqref{G1_actuation}. All non-resonant contributions are in phase with $\uuu_1$. In the Supplemental Material~\cite{Note1} is shown an animated gif-file \qmarks{\texttt{p1\_246\_wide\_G1.gif}} of the full cycle in the time-domain of the line plots in \figref{p1_linePlots}(b) of $p_1$. These animations reveal that the deviation of the analytical solution from the numerical one is  $\sim 1$~\% for $l=2$ at 0.75~MHz, $\sim 15$~\% for $l=4$ at 1.5~MHz, and $\sim 30$~\% for $l=6$ at 2.2~MHz. The deviation increases for increasing mode number, because the wide vertical $G_1$-actuation couples strongly to the vertical 001-mode through the amplitude factor $P^{001}_1$ of \eqref{p_lmn_box_Plmn}, while the frequency-dependency factor $F^{001}(k_0)$ of \eqref{def_Flmn}, which is very small away from resonance, increases significantly as the actuation frequency $f = \cfl k_0$ approaches the vertical-mode resonance $f^{001}_0 = 3.7$~MHz. We can reduce the coupling $P^{001}_1$ to the vertical 001-mode by reducing the width of the $G_1$ actuation profile. This is demonstrated in the animated gif-file \qmarks{\texttt{p1\_246\_narrow\_G1.gif}} of the Supplemental Material~\cite{Note1}, where the width of the $G_1$-actuation has been reduced by a factor of 6 from $\frac12 \Lstar$ to $\frac{1}{12}\Lstar$. This results in a decreased deviation, now less than $\sim 1$~\%, of the analytical result from the numerical one for $p_1$ for all three modes $l = 2,4,6$.

\begin{figure}[t]
\includegraphics[width=\columnwidth]{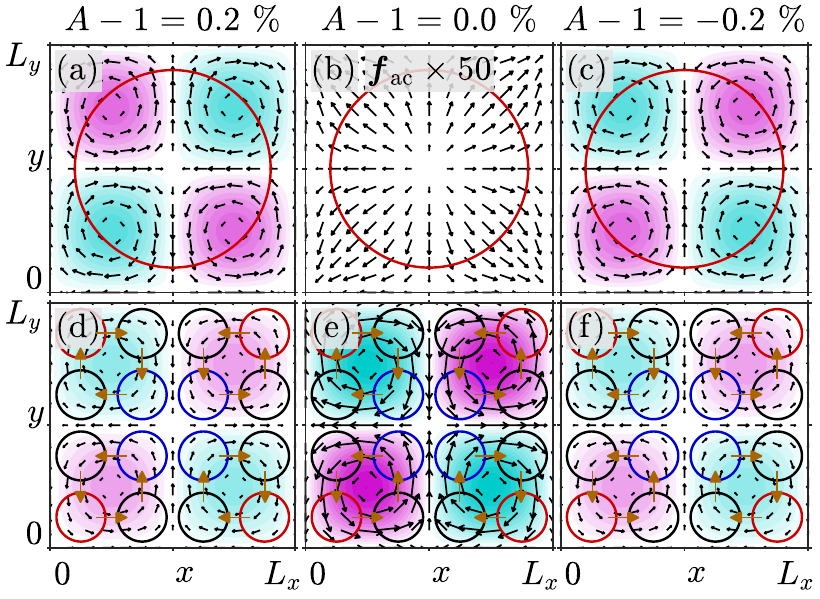}
\caption{\figlab{rotation_direction}  (Color online)
Simulation of the rotating acoustics in the $x$-$y$ plane for the double mode $200+020$ in water for three aspect ratios $A \approx 1.002$, 1, 0.998 each for the non-rotating ($G_1$) and rotating ($G_{16}$) actuation perpendicular to the plane. Color plots between $\pm 10^{-4}\, \SIkg\, \SIm^{-1}\SIs^{-1}$ (light cyan to dark magenta) of the $z$ component $\mathcal{L}_{\mr{ac},z}$ of the angular momentum density $\Lac$, and vector plots of the acoustic body force $\fffac$ (max 2.00 $\SIN\, \SIm^{-3}$). In (a)-(c) we use the single non-rotating Gaussian actuation $G_1$ (large red circles), \eqsref{G1_actuation}{uwall_l00}. In (d)-(f) we use the rotating actuation profile $G_{16}$ of four quadruples of phase shifted Gaussian profiles given in \eqref{G16_all} of \appref{RotAct} (light red [up], black [0], dark blue [down], and black [0] circles, respectively) to generate an actuation rotating in the direction of the thick brown arrows. $A  \approx 1.002$ and 0.998  are chosen to optimize the acoustic rotation for the non-rotating actuation, see \eqref{geom_opt}.}
 \end{figure}

\subsubsection{The acoustic body force and angular momentum density}
\seclab{Resultfac}

In \figref{rotation_direction}, we validate our analysis in \secref{NonRotActuation} of rotating double modes induced by a non-rotating actuation. For the double mode $200+020$, we plot the body force $\fffac=\fff_{020}^{200}$ from \eqref{ALLlmnlmnp}, which according to \eqref{v2_gov} is the source of the bulk-driven acoustic streaming, and the acoustic angular momentum density $\Lac=\calLLL_{020}^{200}$ from \eqref{Llmnlmnp} in the center plane $z=\frac12 \Lz$. We show results for three different cavity aspect ratios $A=\Lx/\Ly$ for both the non-rotating wide Gaussian actuation $G_1(x,y)$, given in \eqref{G1_actuation}, and for the externally-controlled rotating actuation $G_{16}(x,y)$ consisting of four quadruples of narrow Gaussian actuations, see \eqref{G16_all} of \appref{RotAct}. As expected, the non-rotating actuation $G_1$ used in \figref{rotation_direction}(a)-(c) induces no acoustic rotation for a perfect square cavity $A=1$, \figref{rotation_direction}(b), while it induces different rotation directions for $A$ slightly larger and smaller than unity, see respectively \figref{rotation_direction}(a) and (c).  In contrast, the rotating actuation $G_{16}$ used in \figref{rotation_direction}(d)-(f) for the same three aspect ratios, induces an acoustic rotation, which is maximized in the perfect square cavity $A=1$, see \figref{rotation_direction}(e). We show in \eqref{phase_rot} of \appref{RotAct} that this rotating actuation excites the modes $200$ and $020$ with a phase difference $\phiact^{020}-\phiact^{200}\approx 0.5 \pi$ such that the factor $\calF_{020}^{200}$ in \eqref{calF_lmn_phase} is positive, and consequently the rotation pattern in \figref{rotation_direction}(d)-(f) is in the same direction as shown in \figref{rotModes}(c).

In \figref{rot_modes_freq_dep_d} we investigate the strength and rotation direction of the body force $\fff_{020}^{200}$ as a function of the mode separation $\Deltati_{20}$ or equivalently, the aspect ratio $A$. From \eqref{flmnlmnp}, this dependency is quantified through the factor $\calF_{020}^{200}$ which we obtain numerically from the simulation of the double mode $200+020$ in \figref{rotation_direction} by calculating the acoustic angular momentum $\Lac$ and using \eqref{Llmnlmnp},
 \beq{F_202_200_num}
 (\calF_{020}^{200})_\mr{num}\approx \frac{\mathcal{L}_{\mr{ac},z}\big(\frac14 L_x, \frac14 L_y\big)}{\abs{\pa^{200}} \abs{\pa^{020}} \big(\omega \rhofl  c_0^2\big)^{-1}}.
 \eeq
Here, for pressure amplitudes $\abs{\pa^{200}} = \abs{\pa^{020}} = 1$~MPa, the denominator evaluates to $0.954\times10^{-4}\,~\SIkg\,~\SIm^{-1}\SIs^{-1}$ , which is indeed the maximum value found numerically in \figref{rotation_direction}(e), such that $(\calF_{020}^{200})_\mr{num}$ takes the maximum value of unity, as expected. In \figref{rot_modes_freq_dep_d} we plot the largest possible value of $(\calF_{020}^{200})_\mr{num}$ both for the non-rotating actuation used in \figref{rotation_direction}(a)-(c) and for the rotating actuation used in \figref{rotation_direction}(d)-(f). We also plot the analytical expressions for $(\calF_{ml0}^{lm0})_\mr{ana}^\mr{extr}$  derived in \appref{Deltaopt_app} and find a quantitative agreement better than 1~\%.  For the non-rotating actuation (the dark blue curve in \figref{rot_modes_freq_dep_d}), the acoustic rotation direction is determined by the aspect ratio $A$, and it reverses as $A$ crosses over from $A<1$ to $A>1$ around the perfect square $A=1$. %Consequently, for a given aspect ratio, a non-rotating actuation excites a rotation direction of the body force $\fffac$ and thus of the bulk-driven streaming, which is independent of the actuation frequency.
\begin{figure}[t]
\includegraphics[width=\columnwidth]{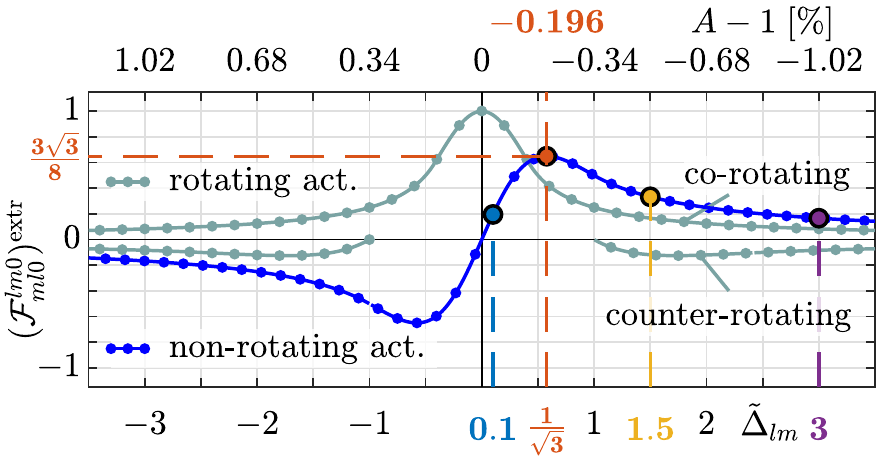}
\caption{\figlab{rot_modes_freq_dep_d}  (Color online)
The largest possible value (positive and negative) of $\calF_{ml0}^{lm0}\propto \fffac$ for varying mode separation $\Deltati_{lm}$ (lower $x$-axis) or aspect ratio $A$ (upper $x$-axis), shown for non-rotating actuations (blue) and rotating actuations (light cyan). The numerical values (dots) are found by evaluating $(\calF_{020}^{200})_\mr{num}$ from \eqref{F_202_200_num} in the simulation of the double mode $200+020$ shown in \figref{rotation_direction}(a)-(c) for the non-rotating and \figref{rotation_direction}(d)-(f) for the rotating actuation. The analytical curves (solid) are from the expressions derived in \eqref{maxrot_val_nonrot} (non-rotating act.), \eqnoref{maxrot_val_rot_pos} (rotating act., co-rotating), and \eqnoref{maxrot_val_rot_neg} (rotating act., counter-rotating) of \appref{Deltaopt_app}. The black-edged circles correspond to those of the same color in \figref{NonRotActuation}(c).}
\end{figure}
 \begin{figure}[t]
\centering
\includegraphics[width=\columnwidth]{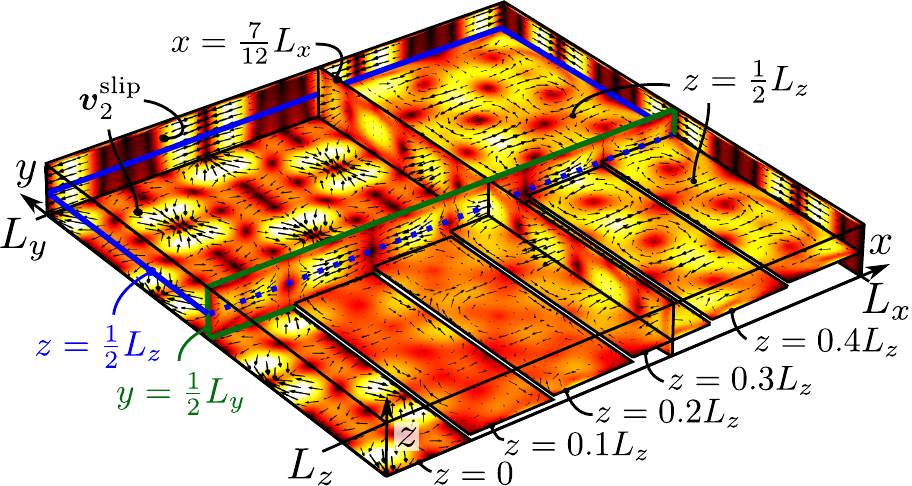}
\caption{\figlab{streaming_3D}  (Color online)
Numerical simulation of the acoustic streaming velocity $\vvv_2$ (black arrows) and its magnitude from 0 (black) to 0.17~mm/s (white) induced by the non-rotating actuation $G_1$. The setup and the parameters are the same as in  \figref{p1_linePlots}  with the parameters listed in \tabsref{fluid_params}{params246}. The horizontal plane at $z=\frac12 \Lz$ (dark blue edge) and the vertical plane at $y=\frac12 \Ly$ (light green edge) are displayed in \figref{streaming246}.}
\end{figure}
For the rotating actuation (the light blue curves in \figref{rot_modes_freq_dep_d}), the rotation direction can be either co- and counter-rotating (at different frequencies) with respect to the rotation direction of the actuation. However, for small mode separations $|\Deltati_{lm}|<1$, which for the double mode $200+020$ corresponds to $|A-1| < 0.34 \%$, the rotation direction of the acoustic field can only be co-rotating. For further details, see \appref{Deltaopt_app}.

%The corresponding analysis for a rotating actuation  is more complicated. For small mode separations $|\Deltati_{lm}|<1$, which for the double mode $200+020$ corresponds to $|A-1| < 0.34 \%$, the rotation direction of the acoustic field is the same as that of the actuation, $(\calF_{ml0}^{lm0})^\mr{extr} > 0$. Remarkably, for larger values of $|\Deltati_{lm}|$ and  $|A-1|$, the acoustic rotation direction can both co-rotating and counter-rotating with respect to the rotation direction of the actuation. These different actuation directions are of course found at different actuation frequencies.

\subsubsection{The acoustic streaming}
\seclab{ResultStreaming}

\begin{figure*}[t!]
\centering
\includegraphics[width=\textwidth]{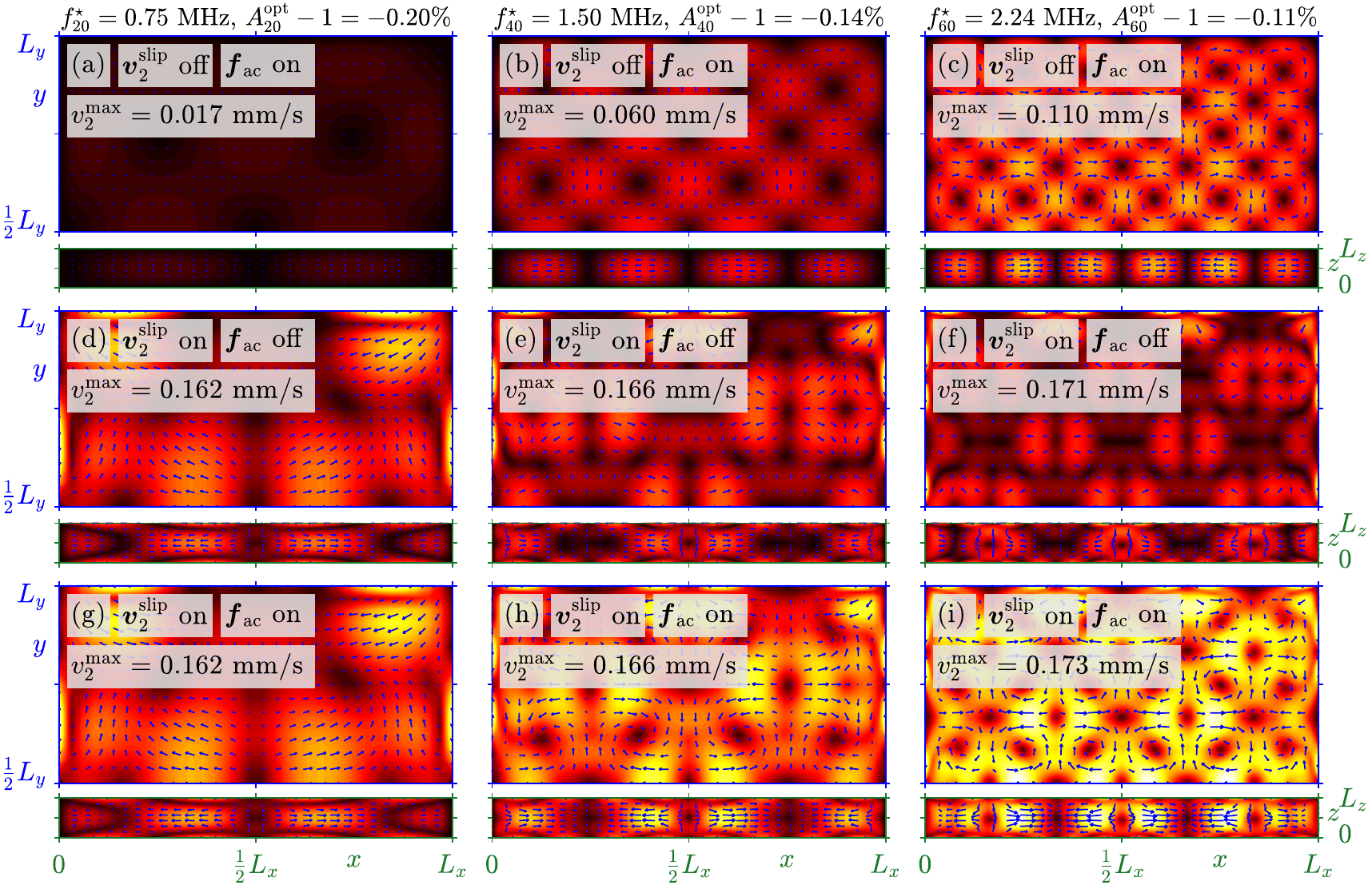}
\caption{\figlab{streaming246}  (Color online)
Simulation of the acoustic streaming $\vvv_2$ in water from 0 (black) to 0.17 mm/s (white) in the horizontal ($x$-$y$) center plane at $z=\frac{1}{2}\Lz$ (dark blue edge as in \figref{streaming_3D}) and the vertical ($x$-$z$) center plane at $y=\frac12 \Ly$ (light green edge as in \figref{streaming_3D}). The used parameters are listed in \tabsref{fluid_params}{params246}. In each column the double mode $l00+0l0$ is excited for $l=2,4,$ and $6$, respectively. The first row (a)-(c) shows the bulk-driven streaming, the second row (d)-(f) shows the boundary-driven streaming, and the third row (g)-(i) shows the total streaming. 3D results corresponding to panel (i) are shown in \figref{streaming_3D}.}
\end{figure*}

In \figref{streaming_3D}, we show the acoustic streaming $\vvv_2$ in 3D resulting from the double mode $600$+$060$ displayed in \figref{p1_linePlots}(a), which is excited by the non-rotating actuation $G_1$. As mentioned above, this situation is similar to that investigated experimentally by H\"agsater \etal~\cite{Hagsater2007}. Remarkably, in the optimized geometry of \figref{streaming_3D}, the horizontal 6-by-6 streaming-roll pattern at any height $0 \leq z \leq \Lz$ is strongest in the center of the cavity, $z=\frac{1}{2}\Lz$, thus indicating its bulk-driven origin. In the bottom plane  $z=0$ (shown in \figref{streaming_3D}) and in the top plane $z=\Lz$ (not shown in  \figref{streaming_3D}), the slip-velocity $\vvv_2^\mathrm{slip}$ is dominated by the gradient term in the expression  \eqnoref{v2_gov_bc_def} for the slip velocity, and has only a weak rotating component.

In \figref{streaming246}, we compare the results for the acoustic streaming $\vvv_2$, see \eqref{v2_gov}, for the double modes $l00+0l0$ with  $l=2,4,6$. The streaming is shown in the horizontal $x$-$y$ center plane at height $z=\frac12\Lz$ (dark blue edge in \figref{streaming_3D}) and in the vertical $x$-$z$ center plane at $y=\frac12\Ly$ (light green edge in \figref{streaming_3D}). In the first row, \figref{streaming246}(a)-(c), we show the bulk-driven streaming alone ($\fffac$ on,  $\vvv_2^{\mr{slip}}$ off); in the second row, \figref{streaming246}(d)-(f), we show the boundary-driven streaming alone ($\fffac$ off,  $\vvv_2^{\mr{slip}}$ on); in the third row, \figref{streaming246}(g)-(i), we show the total streaming ($\fffac$ on,  $\vvv_2^{\mr{slip}}$ on). Comparing the bottom-row and top-row panels in \figref{streaming246} strongly indicate that the respective 4-by-4 and 6-by-6 streaming-roll patterns in Figs. \fignoref{streaming246}(h) and (i) are bulk-driven. Furthermore, \figref{streaming246}(a)-(c) show that the bulk-driven acoustic streaming becomes stronger for higher mode number $l$ and thus for higher frequency $f$. In contrast, the boundary-driven acoustic streaming \figref{streaming246}(d)-(f) stays almost constant in amplitude and shows no frequency dependency. Note that the bulk-driven streaming in \figref{streaming246}(a)-(c) has its maximum velocity $v_2^\mr{max}$ in the bulk of the fluid whereas the boundary-driven streaming in \figref{streaming246}(d)-(f) has its maximum velocity $v_2^\mr{max}$ at the boundaries.

\subsubsection{Comparison with the analytical expression for the bulk-driven streaming}
\seclab{ResultBulkAna}

\vspace*{-3mm}

When the no-slip condition on the vertical side walls is ignored, we can derive an approximate analytical expression for the bulk-driven streaming $(\vvv_{2}^\mr{blk})_{l0}$ for the double modes $l00+0l0$. The details are given in \appref{analytic_solution_v2}, where we find  $(\vvv_{2}^\mr{blk})_{l0}$ to be proportional to the acoustic energy flux density $\SSS_{l0}=\SSS_{0l0}^{l00}$ of \eqref{Slmnlmnp},
 \bal \eqlab{v2_bulk}
 (\vvv_{2}^\mr{blk})_{l0}=\frac{1}{2} \bigg(\frac{4}{3}+\frac{\etaBfl}{\etafl}\bigg)
 \Bigg[1-\frac{\mr{cosh}\big(\frac{2z-\Lz}{\sqrt{2}}k_{l0}^\star\big)}{
 \mr{cosh}\big(\frac{1}{\sqrt{2}}k_{l0}^\star\Lz\big)}\Bigg]\kapfl \SSS_{l0}.
 \eal
Inserting here  $k_{l0}^\star=\frac{l\pi}{\Lstar}$ and expression~\eqnoref{Slmnlmnp} for $\SSS_{l0}$  with $\calF_{l0}=\frac{3\sqrt{3}}{8}$ from \eqref{Flm_opt}, we obtain the maximum value of the bulk-driven streaming velocity in the center plane $z=\frac12\Lz$,
\bal \eqlab{v2_bulk_max}
 (v_{2}^\mr{blk})_{l0}^\mr{max}&=\frac{3\sqrt{3}\Big[1-\mr{sech}\Big(\frac{l\pi \Lz}{\sqrt{2}\Lstar}\Big)\Big]\Big(\frac{4}{3}+\frac{\etaBfl}{\etafl}\Big) \abs{\pa^{l00}}  \abs{\pa^{0l0}}   }{32\rhofl^2\cfl^3}.
\eal
For the three cases shown in \figref{streaming246}(a)-(c), where we have used $\pa^{l00}=\pa^{0l0}=1\,\SIMPa$ and $l=2, 4, 6$, respectively, we calculate from \eqref{v2_bulk_max} the maximum values 0.018, 0.060, and $0.102~\SImm/\SIs$, which deviate less than 7~\% from the corresponding results 0.017, 0.060, and $0.110~\SImm/\SIs$ from the numerical simulation, listed in \figref{streaming246}(a)-(c). For $l=2$, the analytical prediction of $0.018~\SImm/\SIs$ over estimates by 5~\% the numerical value $0.017~\SImm/\SIs$ because it neglects the significant viscous damping from the vertical side walls in this case. For $l=6$, the analytical prediction of $0.102~\SImm/\SIs$ under estimates by 7~\% the numerical value $0.110~\SImm/\SIs$, because the latter includes the coupling to the vertical 001-mode through the wide $\frac12\Lstar$-width vertical $G_1$-actuation as discussed in \secref{ResultPressure}. When changing to the narrow $\frac{1}{12}\Lstar$-width vertical $G_1$-actuation, also discussed in \secref{ResultPressure}, the coupling to the vertical 001-mode diminish so much that the numerical streaming amplitudes becomes 0.017, 0.060, and $0.103~\SImm/\SIs$ for $l= 2,4,6$, respectively, where the latter now deviates less than 1~\% from the analytical prediction $0.102~\SImm/\SIs$. These numerical results for the acoustic streaming is shown in the file \qmarks{\texttt{fig\_08\_narrow\_G1.pdf}} in the Supplemental Material~\cite{Note1}.

\subsubsection{Bulk-driven versus boundary-driven streaming rolls}
\seclab{bulk_vs_boundary}

We address the question on whether the horizontal streaming rolls in the $xy$-plane are due to the bulk-driven streaming or the boundary-driven streaming. We restrict the discussion to horizontal double modes $l00+0l0$, for which we have found the analytical expressions for the bulk-driven streaming $(\vvv_{2}^\mr{blk})_{l0}$, \eqref{v2_bulk}, and the slip velocity $\vvv_2^\mr{slip}$, \eqref{v2_gov_bc_def}, at the top ($z=\Lz$) and bottom ($z=0$) boundaries.

According to \eqref{v2_gov_bc_def}, the boundary-driven streaming velocity at the double-mode resonance $l00+0l0$ is driven by a slip-velocity, which contain the term  $\frac12 \kapfl \SSSac$ that rotates in the direction of the acoustic energy flux density $\SSSac$, just as the bulk-driven streaming, see \eqref{v2_bulk}. The other terms in the slip velocity have similar  magnitude, but they are gradient terms which cannot drive a rotating streaming in the horizontal $xy$-plane. Therefore, as seen in \figref{streaming246}(d)-(f), the boundary-driven streaming does not lead to horizontal flow rolls as clearly as the bulk-driven streaming does, see \figref{streaming246}(a)-(c). There may, however, exist horizontal planes at a few specific heights $z$, where the horizontal flow-roll pattern appears more clearly as suggested by Lei \etal~\cite{Lei2014} and analyzed further by Skov \etal~\cite{Skov2019}.

By using \eqsref{v2_gov_bc_def}{v2_bulk}, we compute the ratio of the horizontally rotating bulk-driven streaming
$\abs{(\vvv_{2}^\mr{blk}(z=\frac{\Lz}{2}))_{l0}}$ in the center of the cavity to the horizontally rotating boundary-driven streaming $\abs{\frac12\kapfl \SSSac(z=0)}$ at the bottom boundary,
 \bal \eqlab{vbu_vslip}
 \frac{\abs{(\vvv_{2}^\mr{blk}(z=\frac{\Lz}{2}))_{l0}}}{
 \abs{\frac12 \kapfl \SSS_{l0}(z=0)}}
 = \bigg(\frac{4}{3}+\frac{\etaBfl}{\etafl}\bigg)\bigg[1-\mr{sech}\Big(\frac{l\pi \Lz}{\sqrt{2}\Lstar}\Big)\bigg].
\eal
This velocity ratio takes values from 0 (boundary-driven streaming dominates) for a flat cavity, $l \Lz \ll \Lstar$, to $\frac{4}{3}+\frac{\etaBfl}{\etafl}$ (bulk-driven streaming dominates) for a high cavity, $l \Lz \gg \Lstar$. For water, the velocity ratio~\eqnoref{vbu_vslip} lies between 0 and 4.13, and in the setup of Figs.~\fignoref{p1_linePlots}-\fignoref{streaming246} with $\Lz=0.1\Lstar$ and water as the fluid, we obtain the values of the velocity ratio to be 0.38, 1.22 and 2.09 for $l=2$, 4 and 6, respectively. This result indicates that, that the 6-by-6-streaming roll pattern with $l=6$ observed by Hags\"ater \textit{et al.}~\cite{Hagsater2007} is predominantly bulk-driven, as also seen by comparing \figref{streaming246}(c) and (i). In contrast, for the lower mode $l=2$, the horizontal streaming rolls are predominantly boundary-driven, which is seen by comparing \figref{streaming246}(d) and (g).

%\begin{widetext}
%\rule{0ex}{3.0ex}
%\begin{figure}[h!]
\begin{figure*}
\centering
\includegraphics[width=\textwidth]{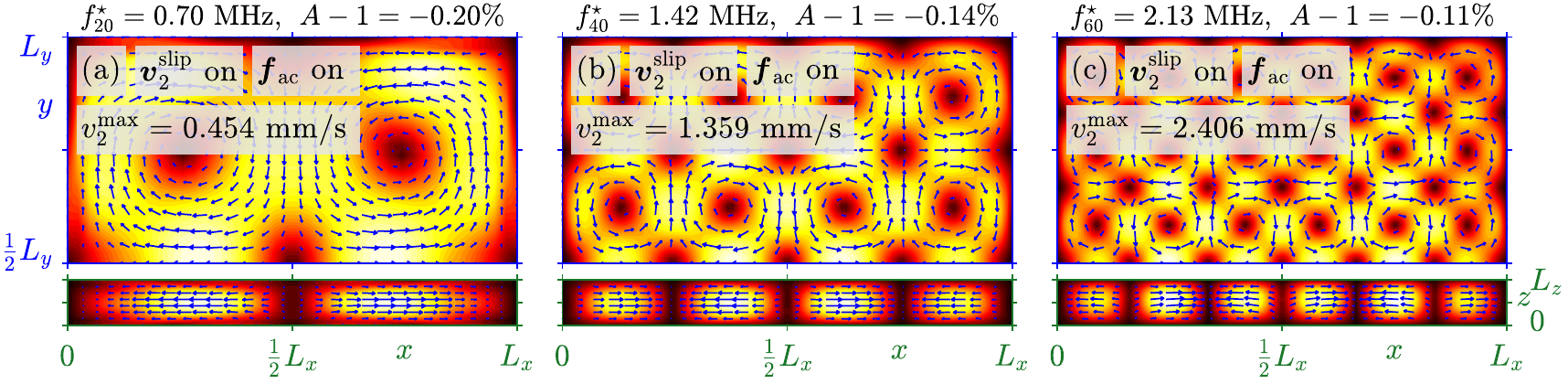}
\caption{\figlab{streaming246_pyridine}  (Color online)
Simulation of the acoustic streaming $\vvv_2$ from 0 mm/s (black) to $v_2^\mr{max}$ (white) for the exact same geometry as in \figref{streaming246}(g)-(i), but with the water replaced by pyridine, having the material parameters listed in \tabref{fluid_params} and actuation amplitude $u_{1z}^0$ and actuation frequency $f_{l0}^\star$ given in \tabref{params246}. The maximum values in (a)-(c) are 2.8, 8.2, and 13.9 times the corresponding values in \figref{streaming246}(g)-(i).}
\end{figure*}
%\end{figure}
%\end{widetext}
% \noindent

The cross-over from boundary- to bulk-driven horizontal streaming rolls may be characterized by the height-to-width ratio $B = l \Lz/\Lstar$: we compute the critical value $B^\mr{crit}$, where the velocity ratio~\eqnoref{vbu_vslip} is unity, and thus,
 \bal \eqlab{Bcrit}
 l\frac{\Lz}{\Lstar} & < B^\mr{crit}, \text{ boundary-driven streaming dominates},
 \nn \\
 l\frac{\Lz}{\Lstar} & > B^\mr{crit}, \text{ bulk-driven streaming dominates}.
 \eal
For water, \eqref{vbu_vslip} leads to the ratio $B^\mr{crit}_\mr{water}=0.35$, and since $\Lz=0.1 \Lstar$ in the setup of Figs.~\fignoref{p1_linePlots}-\fignoref{streaming246}, we find that the horizontal streaming rolls in the double modes $l00+0l0$ with $l \leq 3$ are predominantly boundary-driven, while for $l \geq 4$ they are bulk-driven.

\subsubsection{Enhancement of the bulk-driven streaming velocity by increasing the bulk viscosity}
\seclab{ResultBulkVisc}

For a given resonance mode with a given acoustic pressure amplitude in a given geometry, we obtain from \eqsref{v2_bulk_max}{v2_gov_bc_def} the following scaling laws for the bulk- and boundary-driven streaming, respectively:
 \bsubal{v2_scaling}
 \eqlab{v2_scaling_bulk}
 v_2^\mr{blk} &\propto \frac{1}{\rhofl^{2}\cfl^{3}} \Big(\frac{4}{3}+\frac{\etaBfl}{\etafl}\Big),
 \\
 \eqlab{v2_scaling_slip}
 v_2^\mr{slip} &\propto \frac{1}{\rhofl^{2}\cfl^{3}}.
 \esubal
To study the consequences of these scaling laws, we have chosen to compare water with the heterocyclic organic compound pyridine having the chemical formula C$_5$H$_5$N. \tabref{fluid_params} reveals that pyridine has the same material parameters (within 5~\%) as water, but a bulk viscosity $\etaBfl$ that is $25.1$ times larger than that of water. Using these parameter values in \eqref{v2_scaling}, we find a large enhancement of the bulk-driven streaming, while the boundary-driven streaming is nearly unchanged,
 \bsubalat{v2_scaling_pywa}{2}
 \eqlab{v2_scaling_bulk_pywa}
 v_{2,\mr{pyridine}}^\mr{blk} &= & \; 21.3 &\:v_{2,\mr{water}}^\mr{blk},
 \\
 \eqlab{v2_scaling_slip_pywa}
 v_{2,\mr{pyridine}}^\mr{slip} &= & \;1.2 &\:v_{2,\mr{water}}^\mr{slip}.
 \esubalat
We test this prediction in \figref{streaming246_pyridine}, where we show simulations of the total acoustic streaming in pyridine in the same geometry as in \figref{streaming246}(g)-(i) but with adjusted actuation amplitude $u_{1z}^0$ and frequency $f$ given in \tabref{params246} to maintain the same pressure amplitudes $\pa^{l00}=\pa^{0l0}=1\ \SIMPa$. As expected, we find that the total acoustic streaming for the three double modes is enhanced compared to that in water shown in \figref{streaming246}(g)-(i), and the enhancement factors are 2.8, 8.2, and 13.9, respectively. The increasing factors reflect the increasing weight of bulk-driven relative to boundary-driven streaming as the mode index $l$ increases. The enhancement factors for the bulk-driven streaming alone (not shown in \figref{streaming246_pyridine}) are 21.4, 21.1, and 21.2, as predicted by \eqref{v2_scaling_bulk_pywa},  while they are 1.2, 1.2, and 1.3 for boundary-driven streaming alone (also not shown in \figref{streaming246_pyridine}) in good agreement with \eqref{v2_scaling_slip_pywa}.

For all three cases $l=2$, 4, and 6, the horizontally rotating bulk-driven streaming in the center $z=\frac12 \Lz$ is larger than the horizontally rotating boundary-driven streaming at the top-bottom boundaries, $z=0$ and $z=\Lz$. This is also predicted by \eqref{Bcrit}, since $l\frac{\Lz}{\Lstar}=0.2, 0.4, 0.6 > B^\mr{crit}_\mr{pyridine} = 0.08$ for all three modes in pyridine.

\section{Discussion}
\seclab{Discussion}

In the following, we raise a few discussion points related to the results of the above analysis, which are based on the single- and double mode resonances occurring in a closed box-shaped cavity with narrow viscous boundary layers. We note that the idealized, hard-wall, numerical model for the acoustic streaming is strikingly sensitive to small differences of only $0.1\%$ between the cavity side lengths $\Lx$ and $\Ly$. Our results for the resonance frequency in \eqref{res_wave_numbers} and the optimal aspect ratio in \eqref{A_opt} are therefore crucial in the investigation of the largest possible bulk-driven acoustic streaming. This optimization was not taken into account in the experimental study by Hags\"ater \textit{et al.}~\cite{Hagsater2007}, and the corresponding numerical simulation by Skov \etal~\cite{Skov2019} of the shallow, nearly-square cavity of aspect ratio of $A=0.990$. As this aspect ratio differs from the optimal value $A = 0.999$ for the hard-wall device in \figref{streaming246}(i), it may be possible to increase the bulk-driven acoustic streaming significantly by fine tuning the geometry.

A priori, one may argue that it is difficult to obtain these optimal conditions with a relative precision of the order of  0.1~\% in experiments. However, comparing our \figref{streaming_3D} with the experimental results in Hags\"ater \etal~\cite{Hagsater2007}, we find good qualitative agreement even though none of the optimal conditions were considered when designing the nearly-square fluid cavity. Another example is provided by the straight microchannel with a square cross section and non-rotating actuation that  was studied experimentally by Antfolk \etal~\cite{Antfolk2014}. There, a single large vertical streaming roll similar to \figref{rotModes}(b) was observed experimentally and analyzed numerically in terms of a double-mode excitation.

There may be several reasons for why the bulk-driven acoustic streaming can be observed experimentally in spite of the low 0.1-\%-tolerance level of the aspect ratio suggested by our simplified, hard-wall, micro-cavity analysis in \figref{rot_modes_freq_dep_d}. Firstly, there are other damping mechanisms in a real experiment  \cite{Hahn2015}, besides the bulk and boundary-layer damping considered here, which may broaden the resonance peaks significantly and thereby increase the change for overlapping resonances. Secondly, the double-mode body force $\fff_{\lmnp}^{lmn}$ in \eqref{flmnlmnp}, which is the source of the bulk-driven acoustic streaming, depends on the product $\abs{\pa^{lmn}} \abs{\pa^{\lmnp}}$ of the mode amplitudes, and therefore one can obtain a rotating body force even if the pressure is dominated by one of the modes, say $lmn$, as long as the other mode $\lmnp$ is just weakly represented. Thirdly, in typical experiments, the driving frequency is scanned over a broad span of frequencies, say 100~kHz around 2~MHz,  which increases the possibility for exciting double-mode resonances. Fourthly, unintentional asymmetries and imperfections in the experimental setup may cause modes to overlap locally in some regions at some frequencies and in other regions at other frequencies, rather than the idealized homogeneous
pattern at a single frequency shown in \figref{streaming_3D}. This point was also mentioned by Hags\"ater \etal: \qmarks{\textit{If the frequency is shifted slightly in the vicinity of 2.17 MHz, the same vortex pattern will still be visible, but the strength distribution between the vortices will be altered}}. Moreover, in the 3D simulation by Skov \etal~\cite{Skov2019} including the surrounding solid and the piezo-electric transducer, this inhomogeneous distribution of the streaming was indeed reproduced in their Fig. 2(b2). Revisiting the experiment by Antfolk \etal~\cite{Antfolk2014} of a long straight capillary of square cross section, this effect may also be observed, because it is likely that the aspect ratio is slightly less than unity in some regions of the capillary and slightly above unity in other regions, thereby causing the rotation direction to alternate along the channel.

We have in this work mainly considered horizontal acoustics with $n=0$ half-waves in the vertical $z$-direction. A similar analysis can be done for perpendicular acoustics (with respect to the top and bottom boundaries). In \secsref{ResultPressure}{ResultBulkAna}, we have already briefly discussed the role of the vertical single-mode 001 for the pressure $p_1$ and the streaming $\vvv_2$. It is of course also possible to consider vertical rotating double modes involving, say, a single half-wave in the vertical direction, $lm1+\lp\mp1$. For relatively shallow cavities, a vertical excitation implies a relatively high frequency and a correspondingly increased body force $\fffac$. Furthermore, for perpendicular acoustics where $\pp_\perp v_{1\perp}\approx \div \vvv_1=\ii\omega\kapfl p_1$, the slip-velocity $\vvv_2^\mr{slip}$ on the top and bottom boundaries is dominated by the first and last term in \eqref{v2_gov_bc_def},  $\vvv_2^\mr{slip}\approx \frac12 \kapfl \SSSac+\frac{3}{2\omega}\avr{[\ii\pp_\perp v_{1\perp}]\vvv_{1\parallel}}\approx  -\kapfl \SSSac$. Remarkably, this slip velocity is in the opposite direction of the body force $\fffac\propto \SSSac$, so we expect that the bulk- and boundary-driven streaming will be in opposite directions, and thus the cavity could be designed such these streaming flows tend to cancel. A study of this effect is left for future work.

\section{Conclusion}
\seclab{Conclusion}

We have derived analytical expressions for the single-mode pressure resonances $p_1^{lmn}=\pa^{lmn}F^{lmn}R^{lmn}$ and the corresponding boundary-layer damping coefficients $\Gambl^{lmn}$ for a closed box-shaped cavity with narrow viscous boundary layers, see \secref{single_modes} and \appref{box_app}. Based on these expressions, we have shown that the body force $\fffac$ can drive a strong bulk-driven acoustic streaming, if two overlapping single modes are excited simultaneously at the same frequency, thereby forming a double-mode resonance $lmn+\lmnp$, see \secsref{double_modes}{StrongBodyForce}. In contrast to the conventional wisdom that the bulk-driven streaming can be ignored on the scale of a few acoustic wavelengths, these double modes constitute an important example where the bulk-driven streaming is significant.

We have shown  in \eqref{rotS_intS} that the appearance of bulk-driven streaming in closed microcavities is directly related to the amount of rotation (acoustic angular momentum density) of the underlying acoustic fields. Whereas there is no rotation in a single mode, we have demonstrated in \secsref{PhaseCond}{NonRotActuation} that the double modes can be rotating not only due to an externally-controlled rotating actuation, but, remarkably, also due to  a non-rotating actuation coupled with a weak asymmetry in the geometry of the cavity. Furthermore, in \secref{NonRotActuation} we have derived analytical expressions for the optimal aspect ratios  that maximizes the bulk-driven streaming in a nearly-square cavity with a non-rotating actuation. We found that the bulk-driven acoustic streaming is sensitive to small deviations between the cavity side lengths of only $0.1 \%$.

We have validated the theory by direct numerical simulation in \secref{3DSimulation} of the double modes $200+020$, $400+040$, and $600+060$ in a nearly-square cavity with side lengths $\Lx\approx 2000\,\SIum$ and $\Ly\approx 2000\,\SIum$ and height $\Lz=200\, \SIum$, similar to the device investigated experimentally by Hags\"ater \etal~\cite{Hagsater2007}. Using the analytically-known optimal conditions \eqnoref{geom_opt}, for which the bulk-driven acoustic streaming is maximized, we have avoided the otherwise required time-consuming parametric sweeping in geometry and frequency to locate these optimal conditions. The numerical results shown in Figs.~\fignoref{p1_linePlots}-\fignoref{rot_modes_freq_dep_d} agrees with the analytical expressions with a relative accuracy better than 1~\%.

In \figref{streaming_3D}, the 3D simulation of the acoustic streaming of the double-mode $600+060$ is seen to reproduce the horizontally rotating 6-by-6 streaming-roll pattern observed by Hags\"ater \textit{et al}. In \figref{streaming246} we showed that this streaming pattern at the higher frequency 2.24~MHz is dominated by the bulk-driven streaming, whereas the streaming of the double mode 200+020 at the lower frequency 0.75~MHz is dominated by the boundary-driven streaming. This frequency-dependent crossover occurs because the body force $\fffac$ increases as frequency to the power two, while the slip velocity $\vvv_2^\mr{slip}$ depends only weakly on frequency, see \eqref{v2_gov}. An analytical expression for computing the crossover for the double mode $l00+0l0$ is given in \eqref{vbu_vslip}.

Finally, in agreement with Eckart \cite{Eckart1948}, we have shown in \eqref{v2_bulk} that the bulk-driven acoustic streaming increases with the ratio $\frac{\etaBfl}{\etafl}$ between the bulk and the dynamic viscosity. This dependency on  $\etaBfl$ was studied numerically in  \secref{ResultBulkVisc} by exchanging the water in the microcavity with pyridine. As shown in \figref{streaming246_pyridine}, we found that the bulk-driven streaming for pyridine was enhanced by a factor $\simeq 21$ relative to water. This prediction obviously calls for experimental validation.

In this paper we have pointed out the significance of the bulk-driven acoustic streaming in resonating acoustic cavities even when the geometry length scale is comparable to a few half-wave lengths. The fundamental requirement for obtaining this effect, is the existence of two overlapping resonance modes excited simultaneously at the same frequency. This requirement is easily fulfilled in integrable geometries in 3D, such as rectangular, cylindrical, and spherical cavities, or in 2D, such as rectangular and circular cross sections of long, straight capillary channels, all of which are frequently encountered in experiments reported in the literature. The insight provided by our analysis is therefore relevant and important for acoustofluidics in general, both for fundamental studies and for technological applications.

\appendix

\section{The acoustic pressure in a box with boundary layers}\seclab{box_app}
In this section, we derive the solution for the acoustic pressure $p_1$ with boundary layers in a box-shaped cavity $\Omega$ for which $0\leq x\leq \Lx$, $0\leq y\leq\Ly$, and $0\leq z\leq\Lz$. The governing equation and boundary conditions for $p_1$ is given in \eqref{p1_gov}. We consider first an actuation $U_{1\perp}(x,y)$ which is only non-zero at the bottom boundary $z=0$. The solution for a general actuation on all six boundaries is then finally constructed by superposition. Using separation of variables to solve the Helmholtz equation \eqnoref{p1_gov_helmholtz} gives $p_1$ in the following form,
 \bal\eqlab{solution_form}
 p_1&=X(x)Y(y)Z(z),\quad \kc^2=\kx^2+k_y^2+\kz^2,
 \nn \\
 X(x)&=C_x\cos(\kx x)+S_x\sin(\kx x), \nn \\
 Y(y)&=C_y\cos(\ky y)+S_y\sin(\ky y),\nn \\
 Z(z)&=C_z\cos(\kz z)+S_z\sin(\kz z).
 \eal
Inserting this form into boundary condition~\eqnoref{p1_gov_bc} gives,
 \bsubal{bc_alpha}
 \eqlab{bc_alpha_U}
 \big[\pp_\perp+\alpha_\perp\big]p_1 &= \frac{k_0^2 U_{1 \perp}}{\kapfl},\\
 \eqlab{alpha_definition}
 \alpha_\perp = \alpha_\perp(k_\perp) &= \frac{1+\ii}{2}\delta\:(\kc^2-k_\perp^2),\quad \perp = x,y,z,
 \esubal
where $\perp$ is the \textit{inward} direction perpendicular to the boundary, and the quantity $\alpha_\perp$ is the boundary-layer correction to the \textit{inward}-normal derivative of $p_1$. Note that $\pp_\perp$ has a sign change at two opposite boundaries, say $x=0$ and $x=\Lx$, for which $\pp_\perp=\pp_x$ and $\pp_\perp = -\pp_x$, respectively, whereas $\alpha_\perp(k_\perp)$ depends on $k_\perp^2 = k_x^2$, without a sign change at opposite boundaries. We assume in the following that the length scale introduced by $\alpha_\perp$ is much longer than the cavity length scale $L_\perp$,
 \beq{alpha_assumption}
 |\alpha_\perp|  L_\perp \ll 1, \quad \perp = x,y,z.
 \eeq

According to \eqref{bc_alpha}, the boundary conditions at the stationary boundaries at $x=0$ and $x=\Lx$ are
 \bsubalat{p_bc_x}{2}
 \eqlab{p_bc_x_0}
 &\underline{x=0}: &C_x \alpha_x +S_x\kx & =0,\\
 &\underline{x=L_x}:& \quad C_x[\kx\sin(\kx \Lx)+  \alpha_x\cos(\kx \Lx)]&
 \nn \\
 \eqlab{p_bc_x_Lx}
 &\hspace*{3em}+ &S_x[-\kx\cos(\kx\Lx) +\alpha_x \sin(\kx \Lx)] & =0,
 \esubalat
where $\alpha_\perp = \alpha_x$ in both equations, whereas $\pp_\perp = \pp_x$ in \eqref{p_bc_x_0}, while  $\pp_\perp = -\pp_x$ in \eqref{p_bc_x_Lx}. Non-trivial solutions with $C_x$ and $S_x$ different from zero must satisfy the usual criterion for the equation determinant $\Dx(\kx)$,
\bsubal{determinant}
\eqlab{determinant_cond}
\Dx(\kx) &=0, \text{ with }\\
\eqlab{determinant_def}
\Dx(\kx) &=\big(\kx^2-\alpha_x^2\big) \sin(\kx \Lx)+2\kx \alpha_x\cos(\kx\Lx).
\esubal
We write the solutions $\kx=\kxnx$ to \eqref{determinant_cond} as perturbations away from the inviscid solutions $\Kxnx$,
 \beq{kx_exp}
 \kxnx=\Kxnx+\deltakxnx, \quad \Kxnx=\frac{l\pi}{\Lx}, \quad l=0,1,2, \dots,
 \eeq
where $\deltak_x^l L_x\ll 1$. By inserting this into \eqref{determinant_def}, expanding in $\deltakxnx$, and writing $\alpha^l_x = \alpha_x(\kxnx)$, we find,
\bal
 \Dx(\kxnx) &=(-1)^l\Big\{\big[(\Kxnx+\deltakxnx)^2-(\alphaxnx)^2\big]\deltak_x^l L_x \nn \\
 &\hspace{20mm}+2(\Kxnx+\deltakxnx)\alphaxnx\Big\}.
\eal
From assumption \eqnoref{alpha_assumption},  $(\alphaxnx)^2\deltak_x^l L_x=[\alphaxnx L_x] \deltakxnx\alphaxnx$ is much smaller than  $2\deltakxnx\alphaxnx$, so it can be ignored. We then factorize out $\Kxnx+\deltakxnx=\kxnx$ and obtain,
 \bal
 \Dx(\kxnx) &\approx (-1)^l \kxnx \Big\{(\Kxnx+\deltakxnx)\deltakxnx L_x+2\alphaxnx\Big\}
 \nn \\
 \eqlab{DxEpsK}
 & = (-1)^l \kxnx \Big\{\kxnx(\kxnx-\Kxnx) L_x+2\alphaxnx\Big\}.
 \eal
For $l=0$, we have $\Kx^0 = 0$ and $\kx^0(\kx^0-\Kx^0) = (\kx^0)^2$, while for $l>0$ we have $\kxnx(\kxnx-\Kxnx) \approx \frac12(\kxnx+\Kxnx)(\kxnx-\Kxnx)$ to first order in $\deltakxnx$. In either case, by introducing $\Lxnx$ as
 \bal
 \Lx^l=\frac12(1+\delta_{0l})\Lx,
 \eal
the resulting expression for $\Dx(\kxnx)$ for all $l$ is written as,
 \bal \eqlab{determinant_approx}
 \Dx(\kxnx)\approx (-1)^{l}\kxnx\Lxnx\bigg[ (\kxnx)^2- (\Kxnx)^2+\frac{2\alphaxnx}{\Lxnx}\bigg].
 \eal
The wavenumbers $\kxnx$ thus fulfil the zero-determinant criterion~\eqnoref{determinant_cond} $\Dx(\kxnx)=0$ if
 \beq{kxnx_sqr}
 (\kxnx)^2=(\Kxnx)^2-\frac{2\alphaxnx}{\Lxnx},
 \eeq
and similarly for $(\kyny)^2$ for all $l$ and $m$. From \eqsref{solution_form}{p_bc_x_0} with $\kx=\kxnx$, we obtain the corresponding eigenfunctions $X^l(x)$,
 \beq{X_sol}
  X^{l}(x)=\cos(\kxnx x)-\frac{\alpha_x^{l}}{\kxnx}\sin(\kxnx x),
 \eeq
where the prefactor $C_x^l$ is absorbed into $Z(z)$ in \eqref{solution_form}.
We examine the ratio $\frac{\alpha_x^l}{k_x^l}$ for all $l$ using \eqref{kxnx_sqr},
 \bsubalat{alpha_to_k}{3}
 \bigg(\frac{\alpha_x^0}{k_x^0}\bigg)^2 &= -\frac 12 \Lx \alpha_x^0  &&\ll 1,
 & \qquad &l=0, \\
  \frac{\alpha_x^l}{k_x^l} &= \frac{1}{l\pi}L_x  \alphaxnx &&\ll 1,
  & \qquad &l>0.
 \esubalat
By assumption \eqnoref{alpha_assumption}, we note that even for $l=0$ this ratio is small, but not as small as for $l>0$.

By inserting $\kxnx$ from \eqref{kxnx_sqr} into \eqref{X_sol} and Taylor expanding in $\alphaxnx/\kxnx$, we recover the usual inviscid hard-wall eigenfunctions $\cos\big(\Kxnx x\big)$ plus a small correction of the order $\alpha_x^l L_x$, due to the boundary layers,
 \bal
  X^{l}(x)\approx \cos\big(\Kxnx x\big)+\alpha_x^l L_x
  \begin{cases}
   \frac{x}{\Lx}\big(\frac{x}{\Lx}-1\big), &l=0,  \\[1mm]
  \frac{\sin(\Kxnx x) }{\Kxnx \Lx} \big(\frac{2x}{\Lx}-1\big), &l>0.
  \end{cases}
 \eal
Note that $(\pp_x+\alpha_x^l)X^l|_{x=0}=(\pp_{-x}+\alpha_x^l)X^l|_{x=\Lx}=0$ as required by the boundary condition \eqref{bc_alpha}. Similar expressions are valid for the $y$-eigenfunctions $Y^m(y)$.

In general, the pressure is an infinite sum of the eigenfunctions,
 \bsubal{p_1_sum}
 p_1 &= \sum_{l,m=0}^{\infty}
 X^{l}Y^{m}Z^{lm}(z),\\
 \eqlab{ZlmDef}
 Z^{lm}(z) &= C_z^{lm}\cos(\kz^{lm} z)+S_z^{lm}\sin(\kz^{lm} z),\\
 \eqlab{kzlmDef}
 (k_z^{lm})^2 &= \kc^2-(\kxnx)^2-(\kyny)^2,
 \esubal
where $k_z^{lm}$ depends on the angular frequency $\omega$ of the actuation through $\kc$. At $z=\Lz$, the sets of coefficients $C_z^{lm}$ and $S_z^{lm}$ satisfy a condition similar to \eqref{p_bc_x_Lx}, which leads to
\bal \eqlab{p_bc_z_Lz}
S_z^{lm}=\frac{\kz^{lm}\sin(\kz^{lm} \Lz)+\alpha_z^{lm} \cos(\kz^{lm}\Lz)}{\kz^{lm}\cos(\kz^{lm}\Lz)-\alpha_z^{lm}\sin(\kz^{lm} \Lz)}C_z^{lm},
\eal
where $\alpha_z^{lm} = \alpha_z(k_z^{lm})$. Further, at $z=0$ we have a condition similar to \eqref{p_bc_x_0} with $k_0^2 \kapfl^{-1}U_{1z}(x,y)$ on the right-hand side. Combined with \eqsref{p_1_sum}{p_bc_z_Lz}, the boundary condition at $z=0$ becomes,
 \bal \eqlab{p_bc_z_0}
 \sum_{l,m=0}^{\infty}  \frac{X^{l}(x)Y^{m}(y)\mathcal{D}_z(\kz^{lm}) C^{lm}_z}{\kz^{lm}\cos(\kz^{lm}\Lz)-\alpha_z^{lm} \sin(\kz^{lm} \Lz)}=\frac{k_0^2U_{1z}(x,y)}{\kapfl},
 \eal
where $\D_z$ is defined similar to \eqref{determinant_def}. To find the coefficients $C^{lm}_z$, we write the wall actuation function $U_{1z}(x,y)$ as a generalized Fourier series in the functions $X^l(x)$ and $Y^m(y)$ that form a complete basis set on the interval $0\leq x\leq \Lx$ and $0\leq y\leq\Ly$ to order $\alpha_x^l \Lx$ and $\alpha_y^m\Ly$, respectively,
 \bsubal{app_Fourier_p_conf}
 \eqlab{app_fourier_p_expansion}
 U_{1z}(x,y)&=\sum_{l,m=0}^\infty \Uhat^{lm}_{1z}X^l(x)Y^m(y),
 \\
 \eqlab{Uhat_def}
 \Uhat^{lm}_{1z}&=\frac{\int_0^{\Ly}\int_0^{\Lx}
 U_{1z}(x,y) X^l(x)Y^m(y)\ \mr{d} A}{\int_0^{\Ly}\int_0^{\Lx}
 \big[X^l(x)Y^m(y)\big]^2\ \mr{d} A}.
 \esubal
Inserting the expansion \eqnoref{app_fourier_p_expansion} into \eqref{p_bc_z_0}, we obtain the amplitudes $C_z^{lm}$,
 \beq{Az_sol}
 C_z^{lm}\approx \frac{\kz^{lm}\cos(\kz^{lm}\Lz)-\alpha_z^{lm}\sin(\kz^{lm}\Lz)}{\mathcal{D}_z(\kz^{lm})} \frac{k_0^2 \Uhat_{1z}^{lm}}{\kapfl}.
 \eeq
%.
Finally, using \eqref{p_bc_z_Lz} for $S_z^{lm}$ and \eqref{Az_sol} for $C_z^{lm}$ in \eqref{p_1_sum} for $p_1$, yields the expression for the pressure,
 \bsubal{p1_Zlm_sol_box}
 \eqlab{p1_sol_box}
 &p_1 =  \sum_{lm=0}^{\infty}\frac{k_0^2 k_z^{lm}  \Uhat_{1z}^{lm}}{\kapfl \Dz(\kz^{lm})} X^{l}(x)Y^{m}(y)Z^{lm}(z),\\
 \eqlab{Zlm}
 &Z^{lm}(z) =
 \cos\big[\kz^{lm}(\Lz-z)\big]-\frac{\alpha_z^{lm}}{
 \kz^{lm}}\sin\big[\kz^{lm}(\Lz-z)\big],\\
 \eqlab{kzlmGam}
 &\big[k_z^{lm}(k_0)\big]^2 = (1+\ii\Gamfl)k_0^2-(\kxnx)^2-(\kyny)^2.
 \esubal
The infinite sum in \eqref{p1_sol_box} is general and applies to all frequencies and actuations at the bottom boundary. For a given actuation $U_{1z}$, this solution is largest for frequencies $\omega = \cfl k_0$ where $\Dz(\kz^{lm})$ is smallest, which gives the hard-wall resonances $lmn$ studied below. In real systems, the surrounding solid will have its own resonance properties and therefore, the actuation $U_{1z}$ will depend strongly on the frequency. In this case, the largest value of the prefactor $\frac{k_0^2 k_z^{lm} \Uhat_{1z}^{lm}}{\kapfl \Dz(\kz^{lm})}$ is not necessarily found where $\Dz(\kz^{lm})$ is smallest.

\subsection{Single-mode resonances in a box}
Resonance occurs when $\kO = \frac{1}{\cfl}\omega$ equals one of the values $\kO^{lmn}$ that minimizes $\Dz(\kz^{lm})$ in \eqref{p1_sol_box}. We expect these values of  $\kO^{lmn}$ to be near the inviscid values $K_0^{lmn}$, and write similar to \eqref{kx_exp},
 \bsubal{KO}
 \kO^{lmn}&=\KO^{lmn}+\deltakOlmn,
 \\
 \KO^{lmn}&=\sqrt{(\Kxnx)^2+(\Kyny)^2+(\Kznz)^2},\\
 \Kxnx &= \frac{l \pi}{\Lx}, \quad
 \Kyny = \frac{m\pi}{\Ly}, \quad
 \Kznz = \frac{n\pi}{\Lz},
 \esubal
where $\deltakOlmn$ should be chosen to minimize $\D_z(k_z^{lm})$. The modes $l$ and $m$ in the $x$ and $y$ direction together with the resonance condition $k_0=k_0^{lmn}$ fixes the wavenumber $k_z^{lmn} = k_z^{lm}(k_0^{lmn})$ in  \eqref{kzlmGam}  pertaining to the $z$ direction,
 \beq{kzlmnGam}
 (k_z^{lmn})^2 = (1+\ii\Gamfl)\big(k^{lmn}_0\big)^2-(\kxnx)^2-(\kyny)^2.
 \eeq
Combining this expression with \eqref{kxnx_sqr} for $(\kxnx)^2$ and $(\kyny)^2$, we find,
 \bal\eqlab{kz_nxny}
 (\kz^{lmn})^2 &
 = (\KO^{lmn}+\deltakOlmn)^2(1+\ii\Gamfl^{lmn})
 \nn \\
 &\qquad -\bigg[(\Kxnx)^2-\frac{2\alphaxnx}{\Lxnx}\bigg]
 -\bigg[(\Kyny)^2-\frac{2\alphayny}{\Lyny}\bigg]
 \nn \\
 &\approx
 (\Kznz)^2 \!+\! 2\KO^{lmn}\deltakOlmn
 + \ii\Gamfl^{lmn}(k_0^{lmn})^2
 \nn  \\
 & \qquad + \frac{2\alphaxnx}{\Lxnx}+\frac{2\alphayny}{\Lyny}.
 \eal
To evaluate $\D_z(\kz^{lmn})$ in \eqref{p1_sol_box}, we note that $\kz^{lmn}$ is close to $\Kznz$, so we can use \eqref{determinant_approx} with $(\kxnx)^2-(\Kxnx)^2\rightarrow(\kz^{lmn})^2-(\Kznz)^2$, which is found from \eqref{kz_nxny},
\bal\eqlab{Dz_approx}
\mathcal{D}_z(\kz^{lmn}) \approx (-1)^{n}\Lznz\kz^{lmn}\bigg[&2\KO^{lmn}\deltakOlmn+\ii\Gamfl^{lmn}(k_0^{lmn})^2\nn\\
&+\frac{2\alpha_x^{l}}{\Lx^{l}}+\frac{2\alpha_y^{m}}{\Ly^{m}}+\frac{2\alpha_z^{n}}{\Lz^{n}}\bigg].
\eal
The (real) value of $\deltakOlmn$, which minimizes this expression and leads to the resonance wavenumber $k_0^{lmn}$, is then,
\bsubal{Delta_Gamma_xyz} \eqlab{eps0lmn}
 \deltakOlmn &=-\Re\bigg[\frac{\alpha_x^{l}}{\KO^{lmn}\Lx^{l}}+\frac{\alpha_y^{m}}{\KO^{lmn}\Ly^{m}}+\frac{\alpha_z^{n}}{\KO^{lmn}\Lz^{n}} \bigg] \nn \\
 &=-\frac12\KO^{lmn}\Gambl^{lmn},\\
 \Gambl^{lmn}&=\frac{1}{(\KO^{lmn})^2}\Re\bigg[\frac{2\alpha_x^{l}}{\Lx^{l}}+\frac{2\alpha_y^{m}}{\Ly^{m}}+\frac{2\alpha_z^{n}}{\Lz^{n}} \bigg] \nn  \\&=
 \bigg(\frac{\Kxnx}{\KO^{lmn}}\bigg)^2\bigg(\frac{\delta}{\Lyny}+\frac{\delta}{\Lznz}\bigg) \nn\\
 &+ \bigg(\frac{\Kyny}{\KO^{lmn}}\bigg)^2\bigg(\frac{\delta}{\Lznz}+\frac{\delta}{\Lxnx}\bigg)\nn \\
  &+\bigg(\frac{\Kznz}{\KO^{lmn}}\bigg)^2\bigg(\frac{\delta}{\Lxnx}+\frac{\delta}{\Lyny}\bigg),
\esubal
where we have inserted the expressions for $\alpha_x^l$, $\alpha_y^m$, and $\alpha_z^n$ defined by \eqref{alpha_definition}, and introduced the boundary layer damping coefficients $\Gambl^{lmn}$.

We are now in a position to determine the third and last wavenumber $\kz^{lmn}$, the one in the actuation direction. Using the value of $\deltakOlmn$ from \eqref{eps0lmn} in expression \eqnoref{kz_nxny} for $\kz^{lmn}$, we find,
 \bal \eqlab{kzlm_expand}
 (\kz^{lmn})^2&=(\Kznz)^2+\ii \Gamfl^{lmn}(k_0^{lmn})^2\nn \\
 &\hspace{3mm} +\ii \Re\bigg[\frac{2\alphaxnx}{\Lxnx}+\frac{2\alphayny}{\Lyny}\bigg]
 -\Re\bigg[\frac{2\alpha_z^n}{\Lznz}\bigg] \nn \\
 &=(\Kznz)^2+\ii \Gamfl^{lmn}(k_0^{lmn})^2 +\ii (K_0^{lmn})^2 \Gambl^{lmn}-\frac{2\alphaznz}{\Lznz} \nn \\
 &\approx (\Kznz)^2-\frac{2\alphaznz}{\Lznz}+\ii (k_0^{lmn})^2 \Gamma^{lmn} \nn \\
 &=(\kznz)^2+\ii (k_0^{lmn})^2 \Gamma^{lmn},
 \eal
with $\kznz$ defined similar to $\kxnx$ in \eqref{kxnx_sqr} and the total damping coefficient given by,
\bal \eqlab{app_Gamma}
 \Gamma^{lmn}=\Gambl^{lmn}+\Gamfl^{lmn}.
\eal
Note that in comparison with $\kxnx$ and $\kyny$, the wavenumber $\kz^{lmn}$ has an additional dependency on $\Gamma^{lmn}$. The corresponding $z$-dependent functions $Z^{lmn}(z)$ are computed to lowest order in the small parameters by inserting \eqref{kzlm_expand} into \eqref{Zlm},
 \bsub\eqlab{Zlm_res}
 \bal
 Z^{lmn}(z)&\approx (-1)^n \Big[Z^n(z)+ \ii (k_0^{lmn}\Lz)^2 \Gamma^{lmn} Z^n_\mr{act}(z)\Big], \\
 Z^n_\mr{act}(z) &= \begin{cases}
    -\frac12 \big(1-\frac{z}{\Lz}\big)^2, &n=0, \vspace{1mm} \\
  \phantom{-}\frac12 \frac{\sin\big(\Kznz z\big) }{\Kznz \Lz}\big(1-\frac{z}{\Lz}\big) , &n>0.
 \end{cases}
 \eal
 \esub
Here,  $Z^{lmn}(z)$ is close to the eigenfunction $Z^n(z)$ for stationary boundaries, analogous to $X^l(x)$ and $Y^m(y)$, see \eqref{X_sol}, but it contains an extra term, which satisfies the boundary condition at the moving actuated boundary $z=0$.

Finally, we evaluate $\D_z(\kz^{lmn})$ in \eqref{Dz_approx} at resonance $k_0=k_0^{lmn}$ where $\eps_0^{lmn}$ takes the value given in \eqref{eps0lmn},
 \bal \eqlab{Dz_res}
 \D_z(\kz^{lmn})\approx (-1)^{n}\Lznz\kz^{lmn} \ii\Gamma^{lmn} (k_0^{lmn})^2 .
 \eal
and by inserting \eqsref{Dz_res}{Zlm_res} into \eqref{p1_sol_box}, we find an expression for each resonance mode $p_1^{lmn}$,
 \bal \eqlab{p1_lmn_alpha_correct}
 &p_1^{lmn} =
 \frac{\big(k^{lmn}_0\big)^2 \kz^{lmn}  \Uhat_{1z}^{lm}}{\kapfl \Dz(\kz^{lmn})} X^{l}(x)Y^{m}(y)Z^{lmn}(z)
 \\ \nn
 &= \frac{\big(k^{lmn}_0\big)^2\Uhat_{1z}^{lm} }{\kapfl\Lznz} X^l(x)Y^m(y)
 \bigg[ \frac{Z^n(z)}{\ii(k_0^{lmn})^2\Gamma^{lmn}}+\Lz^2 Z_\mr{act}^n(z) \bigg].
 \eal
Note that for any $l$, $m$, and $n$, the product term $X^l(x)Y^m(y)Z^n(z)$ only satisfies the boundary condition \eqref{bc_alpha_U} for a \textit{stationary} boundary $U_{1\perp} = 0$. The actual boundary condition is satisfied by the infinite sum~\eqnoref{p1_sol_box} over terms  $X^l(x)Y^m(y)Z_\mr{act}^n(z)$, because $\pp_zZ_\mr{act}^n=\frac{\Lznz}{(\Lz)^2}$ at $z=0$ and $\pp_zZ_\mr{act}^n=0$ at $z=\Lz$.

As \eqref{p1_lmn_alpha_correct} is only valid exactly at resonance $k_0=k_0^{lmn}$, it must be corrected to deal with frequencies  $k_0\approx k_0^{lmn}$. This is done by adding  $k_0-k_0^{lmn}$ to the right-hand side of \eqref{KO}. By \eqref{Dz_approx}, this procedure is seen to be equivalent to the substitution
\bal \eqlab{substituting}
\ii(k_0^{lmn})^2\Gamma^{lmn}\rightarrow 2 k_0^{lmn}(k_0-k_0^{lmn})+ \ii(k_0^{lmn})^2\Gamma^{lmn}
\eal
in \eqref{p1_lmn_alpha_correct}.

\subsection{Approximate solutions near resonance for actuation at all boundaries}
At resonance, the expression \eqnoref{p1_lmn_alpha_correct} satisfies all boundary conditions to first order in $\alpha_x^l L_x, \alpha_y^m L_y,$ and $\alpha_z^n L_z$. Ignoring these first-order corrections as well as the small $Z^n_\mr{act}$-term in \eqref{p1_lmn_alpha_correct}, the eigenfunctions are approximately equal to the usual hard-wall eigenfunctions $R^{lmn}$,
\bal
R^{lmn}(x,y,z)=\cos(\Kxnx x)\cos(\Kyny y)\cos(\Kznz z).
\eal
To generalize our results from an actuation acting on only the bottom boundary $z=0$ to all six boundaries, we note first that $\Lznz\approx \int_0^{\Lz} [Z^n(z))]^2\, \dd z$, whereby \eqref{Uhat_def} can be rewritten as
 \bal
 \frac{\Uhat_{1z}^{lm} }{\Lznz}&=\frac{\int_0^{\Ly}\int_0^{\Lx}
 U_{1z}(x,y)X^l(x)Y^m(y)\ \mr{d} A}{\Lznz\int_0^{\Ly}\int_0^{\Lx}
 [X^l(x)Y^m(y)]^2\ \mr{d} A} \nn \\
 &\approx \frac{\int_0^{\Ly}\int_0^{\Lx} U_{1z}(x,y) R^{lmn}(x,y,0)\ \mr{d} A}{
 \int_0^{\Lz}\int_0^{\Ly}\int_0^{\Lx} [R^{lmn}(x,y,z)]^2\ \mr{d} V}.
 \eal
Using this expression together with the substitution~\eqnoref{substituting} valid for $k_0\approx k_0^{lmn}$, we obtain the following expression for the resonance modes $lmn$ in a box $\Omega$ with viscous boundary layers and an \textit{inward} normal displacement $U_{1\perp}(\rrr)$ specified on all six boundaries $\pp\Omega$, \eqref{p1_gov_W},
\bsubal{app_p1_sol_general}
p_1^{lmn} &=\pa^{lmn}F^{lmn}(k_0)R^{lmn}(\rrr),\\
\pa^{lmn} &= \frac{\rhofl \cfl^2}{\Gamma^{lmn}}\,  \frac{\int_{\pp\Omega}
 U_{1\perp} R^{lmn}\ \mr{d} A}{\int_\Omega (R^{lmn})^2\, \dd V },\\
 F^{lmn}(k_0)&= \frac{ \frac12 k_0^{lmn}\Gamma^{lmn}}{(\kO-\kO^{lmn})+\ii\frac12 k_0^{lmn}\Gamma^{lmn}}.
\esubal
The quality factor $Q^{lmn}$ for resonance mode $lmn$ is obtained from the linewidth of the acoustic energy density $\Eac$ in \eqref{S}, and since  $\Eac \propto |p_1^{lmn}|^2 \propto |F^{lmn}|^2$, we find
 \bal
 Q^{lmn}
 \approx \frac{1}{\Gamma^{lmn}},
 \eal
where $\Gamma^{lmn}$ is given in \eqref{app_Gamma}.

\section{Maximum acoustic rotation}
\seclab{Deltaopt_app}

In the following, we evaluate the frequency-dependent factor $\calF_{ml0}^{lm0}$ in \eqref{Flmnlmnp} for rotating and non-rotating actuations of the nearly-square cavity considered in \secref{NonRotActuation}. For convenience, we first introduce the double-mode quantities, $\xiti_{lm}$ and $\phi_{lm}^\mr{act}$,
 \bal
 \xiti_{lm}=\kti_0-\kti_{lm}^\star, \qquad \phi_{lm}^\mr{act}=\phiact^{ml0}-\phiact^{lm0},
 \eal
where $\kti_0$ and $\kti_{lm}^\star$ are defined in \eqref{ktilde_def}, and $\phi_{lm}^\mr{act}$ is the difference in the phases by which the modes $lm0$ and $ml0$ are actuated, see \eqref{p_lmn_box_Plmn}. By using these quantities and the expressions for $F^{lm0}$ and $F^{ml0}$ in \eqref{Flmn_normalized}, we calculate the frequency-dependency $\calF_{ml0}^{lm0}$ from \eqref{Flmnlmnp},
 \bal\eqlab{calF_xi_general}
 \calF_{ml0}^{lm0}&=\Re\Big\{\ii \ee^{\ii \phi_{lm}^\mr{act}} F^{lm0}\big[F^{ml0}\big]^*\Big\}
 \\ \nn
 &=\frac{\cos(\phi_{lm}^\mr{act}) 2\Deltati_{lm}+\sin(\phi_{lm}^\mr{act})[1+(2\xiti_{lm})^2-\Deltati_{lm}^2]}{
 [(2\xiti_{lm}-\Deltati_{lm})^2+1][(2\xiti_{lm}+\Deltati_{lm})^2+1]}.
 \eal
This equation expresses the magnitude of the acoustic rotation as a function of the actuation phase difference $\phi_{lm}^\mr{act}$, the actuation frequency $\xiti_{lm}$, and the mode separation $\Deltati_{lm}$, which by \eqref{d_to_k_perturb} is related to the aspect ratio $A$. In the following, we study the non-rotating actuation with $\phi_{lm}^\mr{act}=0$ and the  rotating actuation with $\phi_{lm}^\mr{act}=\pm \frac12 \pi$, and we determine the optimal mode separation $\Deltati_{lm}^\mr{opt}$ and frequency $\xiti_{lm}^\mr{opt}$, for which $\calF_{ml0}^{lm0}$ takes its largest positive or negative value $(\calF_{ml0}^{lm0})^\mr{extr}$.

\subsection{Extremum values of $\calF_{ml0}^{lm0}$ for a non-rotating actuation $\phi_{lm}^\mr{act}=0$}

For a non-rotating actuation, we have $\phi_{lm}^\mr{act}=0$, and \eqref{calF_xi_general} becomes
 \beq{calF_nonrot_xi}
 \calF_{ml0}^{lm0}=\frac{2\Deltati_{lm}}{[(2\xiti_{lm}-\Deltati_{lm})^2+1][(2\xiti_{lm}+\Deltati_{lm})^2+1]}.
 \eeq
The extremum values of this expressions are
 \bsub
 \bal \eqlab{maxrot_val_nonrot}
 (\mathcal{F}_{ml0}^{lm0})^\mr{extr} &= \begin{cases}
 \frac{2\Deltati_{lm}}{(\Deltati_{lm}^2+1)^2}, & \;   |\Deltati_{lm}|< 1,\\[2mm]
 \frac{1}{2\Deltati_{lm}}, & \; |\Deltati_{lm}|> 1,
 \end{cases}
 \eal
which are obtained at the frequencies $\xiti_{lm}^\mr{extr}$ given by
 \bal \eqlab{maxrot_k0_nonrot}
 \xiti_{lm}^\mr{extr}&=\begin{cases}
 0, & \; |\tilde{\Delta}_{lm}|< 1, \\
 \pm\frac12\sqrt{\tilde{\Delta}_{lm}^2-1}, & \; |\tilde{\Delta}_{lm}|> 1.
 \end{cases}
 \eal
 \esub
In \figref{rot_modes_freq_dep_d} (\qmarks{non-rotating act.}) we plot the value $(\mathcal{F}_{ml0}^{lm0})^\mr{extr}$ from this expression as a function of $\Deltati_{lm}$. The optimal value $(\calF_{ml0}^{lm0})^\mr{opt}$ of $(\mathcal{F}_{ml0}^{lm0})^\mr{extr}$ is found at the optimal mode separation $\Deltati_{lm}^\mr{opt}$, and optimal rescaled frequency $\xiti_{lm}^\mr{opt}$,
 \bal \eqlab{Deltati_opt_appendix}
 \xiti^\mr{opt}_{lm}=0, \qquad \Deltati_{lm}^\mr{opt}=\pm \frac{1}{\sqrt{3}}, \qquad  (\calF_{ml0}^{lm0})^\mr{opt}=\pm \frac{3\sqrt{3}}{8}.
 \eal

\subsection{Extremum values of $\calF_{ml0}^{lm0}$ for a rotating actuation $\phi_{lm}^\mr{act}=\pm \frac12 \pi$}

For an externally controlled rotating actuation, we have $\phi_{lm}^\mr{act}=\pm \frac12 \pi$ and \eqref{calF_xi_general} becomes,
 \bsubal{calF_sgnphi}
 \eqlab{calF_rot_xi}
 \calF_{ml0}^{lm0} &= \frac{\big[1+(2\xiti_{lm})^2-\Deltati_{lm}^2\big]\mr{sgn}_\phi^{lm}}{
 [(2\xiti_{lm}-\Deltati_{lm})^2+1][(2\xiti_{lm}+\Deltati_{lm})^2+1]},
 \\
 \eqlab{sgnphi}
 \mr{sgn}_\phi^{lm} &= \sign(\phi_{lm}^\mr{act}).
 \esubal
This expression is more complicated than \eqref{calF_nonrot_xi}, since it only has one extremum for a sufficiently small mode separation $\Deltati_{lm}$, namely where the angular momentum and the actuation are co-rotating, $\sign(\calF_{ml0}^{lm0}) = \mr{sgn}_\phi^{lm}$. For larger values of $\Deltati_{lm}$, the frequency can be tuned to two different extremum values, namely both the co- and counter-rotating cases, $\sign(\calF_{ml0}^{lm0}) = \pm\mr{sgn}_\phi^{lm}$. We therefore identify two branches of extremum values. The co-rotating branch $(\calF_{ml0}^{lm0})^\mr{extr}_\mr{co}$ with subscript \qmarks{co} and the counter-rotating branch $(\calF_{ml0}^{lm0})^\mr{extr}_\mr{cntr}$ with subscript \qmarks{cntr}.

For the co-rotating branch, $(\calF_{ml0}^{lm0})^\mr{extr}_\mr{co}$ is
 \bsub
 \bal \eqlab{maxrot_val_rot_pos}
 (\calF_{ml0}^{lm0})^\mr{extr}_\mr{co} = \mr{sgn}_\phi^{lm} \begin{cases}
 \frac{1-\Deltati_{lm}^2}{(\Deltati_{lm}^2+1)^2}, & \; |\tilde{\Delta}_{lm}|\leq \sqrt{2}-1,\\[2mm]
 \frac{1}{4|\Deltati_{lm}|}, &\;   |\tilde{\Delta}_{lm}|\geq \sqrt{2}-1,
 \end{cases}
 \eal
found at the frequencies $\xiti_{lm}=\xiti_{lm,\mr{co}}^\mr{extr}$,
 \bal \eqlab{maxrot_k0_rot_pos}
 \xiti_{lm,\mr{co}}^\mr{extr} = \begin{cases}
 0, & \;  |\tilde{\Delta}_{lm}| \leq \sqrt{2}-1, \\
 \pm\frac12\sqrt{\tilde{\Delta}_{lm}^2+2\Deltati_{lm}-1}, & \; |\tilde{\Delta}_{lm} |\geq \sqrt{2}-1.
 \end{cases}
\eal
 \esub
For the counter-rotating branch, $(\calF_{ml0}^{lm0})^\mr{extr}_\mr{cntr}$ is
 \bsub
  \bal \eqlab{maxrot_val_rot_neg}
 (\calF_{ml0}^{lm0})^\mr{extr}_\mr{cntr} = -\mr{sgn}_\phi^{lm} \begin{cases}
 \frac{\Deltati_{lm}^2-1}{(\Deltati_{lm}^2+1)^2}, & 1\leq |\tilde{\Delta}_{lm}|\leq \sqrt{2}+1,
 \\[2mm]
 \frac{1}{4|\Deltati_{lm}|}, & |\tilde{\Delta}_{lm}|\geq\sqrt{2}+1,
 \end{cases}
 \eal
found at the frequencies $\xiti_{lm}=\xiti_{lm,\mr{cntr}}^\mr{extr}$,
 \bal \eqlab{maxrot_k0_rot_neg}
 \xiti_{lm,\mr{cntr}}^\mr{extr} = \begin{cases}
 0,& 1\leq |\tilde{\Delta}_{lm}|\leq \sqrt{2}+1, \\[2mm]
 \pm \frac12\sqrt{\tilde{\Delta}_{lm}^2\!-\!2\Deltati_{lm}\!-\!1},& |\tilde{\Delta}_{lm}|\geq \sqrt{2}+1.
 \end{cases}
 \eal
 \esub
In \figref{rot_modes_freq_dep_d} (Rotating act.), we plot for $\phi_{lm}^\mr{act} = +\frac12 \pi$ both the co- and counter rotating branches, $(\calF_{ml0}^{lm0})^\mr{extr}_\mr{co}$ and $(\calF_{ml0}^{lm0})^\mr{extr}_\mr{cntr}$.

From \eqsref{maxrot_val_rot_pos}{maxrot_val_rot_neg}, we find that the largest value $(\calF_{ml0}^{lm0})^\mr{opt}$ of $(\calF_{ml0}^{lm0})$ is the co-rotating double mode found at the optimal mode separation $\Deltati_{lm}^\mr{opt}$, and the rescaled frequency $\xiti_{lm}^\mr{opt}$,
 \bal \eqlab{Deltati_opt_rot_appendix}
 \xiti^\mr{opt}_{lm}=0, \qquad \Deltati_{lm}^\mr{opt}=0, \qquad  (\calF_{ml0}^{lm0})^\mr{opt}= \mr{sgn}^{lm}_\phi.
 \eal
We see that in contrast to the non-rotating actuation~\eqnoref{Deltati_opt_appendix}, the rotating actuation optimizes the acoustic rotation for zero mode separation, $\Deltati_{lm}^\mr{opt}=0$.

\section{The rotating actuation \textit{G}$\mathbf{_{16}}$ of the double mode 200+020} \seclab{RotAct}
In the simulation shown in \figref{rotation_direction}(d-f) of \secref{Resultfac} we excite the double mode $200+020$ by using the rotating actuation,
 \bsub
 \eqlab{G16_all}
 \bal \eqlab{act16}
 u_{1z}=u_{1z}^0 G_{16}(x,y),\quad 0<x<L_x, 0<y<L_y.
 \eal
Here, $G_{16}(x,y)$ is constructed as the sum of 16 narrow Gaussians $G^\odot_1$ as follows. The primary narrow Gaussian
$G^\odot_1(x,y)$ is centered around $(\frac18L_x,\frac18 L_y)$ in the lower left corner of the domain,
 \beq{Gn1def}
 G^\odot_1(x,y)  =
 \exp\Bigg[-\frac{\big(x-\frac{L_x}{8}\big)^2}{\big(\frac{L_x}{16}\big)^2}
 -\frac{\big(y-\frac{L_y}{8}\big)^2}{\big(\frac{L_y}{16}\big)^2}\Bigg].
 \eeq
Four versions of $G^\odot_1$, centered at $(\frac{2\pm1}{8} L_x, \frac{2\pm1}{8} L_y)$ and multiplied by specific phase factors in order to create a positive rotation direction in the $x$-$y$ plane, are added to form a quadruple $G_4$ centered around $(\frac14L_x,\frac14 L_y)$,
 \bal\eqlab{G4}
 &G_4(x,y)  =
 \ee^{-\ii\frac{0\pi}{2}} G^\odot_1(x,y)
 +  \ee^{-\ii\frac{1\pi}{2}} G^\odot_1(x,y-\frac{1}{4}L_y)
 \nn \\
 & +  \ee^{-\ii\frac{2\pi}{2}} G^\odot_1(x-\frac{1}{4}L_x,y-\frac{1}{4}L_y)
 +  \ee^{-\ii\frac{3\pi}{2}} G^\odot_1(x-\frac{1}{4}L_x,y).
 \eal
Finally, by mirroring the quadruple $G_4(x,y)$ across the center lines $x = \frac12 L_x$ and $y=\frac12 L_y$, $G_{16}$ is formed by adding the four resulting quadruples,
 \bal
 \eqlab{G16}
 G_{16}(x,y) &= G_4(x,y) + G_4(\Lx-x,y) \\
 &\quad + G_4(\Lx-x,\Ly-y)\nn + G_4(x,\Ly-y).
 \eal
 \esub
To obtain the complex pressure amplitudes $P_1^{200}$ and $P_1^{020}$ resulting from $G_{16}$, we first evaluate the integral in the denominator of \eqref{uwall_l00}, where $G_1$ is substituted by $G_{16}$,
 \bsubal{G16int}
\int_0^{\Lx}\int_0^{\Ly} G_{16}(x,y)R^{200} \,\frac{\dd x}{\Lx} \,\frac{\dd y}{\Ly} & = 0.0942\times \ee^{\ii0.250\pi},\\
\int_0^{\Lx}\int_0^{\Ly} G_{16}(x,y) R^{020} \,\frac{\dd x}{\Lx} \,\frac{\dd y}{\Ly}  & = 0.0942\times \ee^{-\ii0.250\pi}.
 \esubal
We then use \eqref{uwall_l00} to compute the desired amplitudes,
 \bsubal{pa_rotating_actuation}
P_1^{200}& = \frac{2u_{1z}^0}{\Lz(\Gamfl^{200}+\Gambl^{200}) \kapfl}0.0942\times \ee^{\ii0.250\pi},\\
P_1^{020}& = \frac{2u_{1z}^0}{\Lz(\Gamfl^{200}+\Gambl^{200}) \kapfl}0.0942\times \ee^{-\ii0.250\pi}.
 \esubal
We see that the actuation in \eqref{act16} excites the two modes $200$ and $020$ with a phase difference of
 \beq{phase_rot}
 \phi_\mr{act}^{020}-\phi_\mr{act}^{200} = 0.500\pi.
 \eeq
Finally, we use \eqref{pa_rotating_actuation} to choose the amplitude $u_{1z}^0$ of the actuation such that $|P_1^{200}|=|P_1^{020}|=1$ MPa,
 \beq{uz_rotating_actuation}
 u_{1z}^0=\dfrac{\Lz(\Gamfl^{200}+\Gambl^{200}) \kapfl | P_1^{200}|}{2\times 0.0942} = 1.614\, \SInm,
 \eeq
where we used the parameters for water given in \tabref{fluid_params}.

\section{Approximate solution for the bulk-driven streaming from double modes $l00+0l0$} \seclab{analytic_solution_v2}

In this section, we calculate an approximate expression for the bulk-driven acoustic streaming velocity $(\vvv_{2}^\mr{blk})_{0l0}^{l00}=(\vvv_{2}^\mr{blk})_{l0}$ driven by the acoustic body force $\fffac = \fff_{0l0}^{l00}=\fff_{l0}$ resulting from the combination of the two perpendicular, horizontal modes $l00$ and $0l0$ in a square cavity. As in \eqref{non-rotation_calF}, the double subscript $l0$ refers to the double mode $l00+0l0$. The bulk-driven streaming velocity $(\vvv_{2}^\mr{blk})_{l0}$ satisfies an equation similar to \eqref{v2_gov} except for having no-slip at the boundaries,
 \bsublab{v2_gov_bulk}
 \bal
 \eqlab{v2_gov_cont_bulk}
 0 &= \div(\vvv_{2}^\mr{blk})_{l0}, &&\rrr\in \Omega,
 \\
 \eqlab{v2_gov_navier_bulk}
 \grad (p_{2}^{\mr{blk}})_{l0} &=
  \etafl\lap(\vvv_{2}^\mr{blk})_{l0}
 +\frac{\Gamfl\omega}{\cfl^2}\SSS_{l0},  \hspace{-1mm}&&\rrr\in \Omega,\\
 \eqlab{v2_gov_bc_bulk}
(\vvv_{2}^\mr{blk})_{l0}&= \zerovec,  \qquad &&\rrr\in \pp \Omega.
\eal
\esublab
Here, $\SSS_{l0}$ is defined in \eqref{Slmnlmnp} combined with \eqref{non-rotation_calF} for $\calF_{l0}(k_0)$, \eqref{Klmnlmnp} for $\calKKK_{l0}(\rrr)$, and \eqref{def_Rlmn} for $R_{l0}(\rrr)$,
 \bsubal{SRF_l00_0l0}
 \eqlab{S_l00_0l0}
 \SSS_{l0}(\rrr) &=\frac12 \frac{\big| \pa^{l00}\big|\big|\pa^{0l0} \big|}{\rhofl \omega}  \calF_{l0}(k_0) \calKKK_{l0}(\rrr),
 \\
 \calF_{l0}(k_0)&=\Re\Big\{\ii \Big[ F^{l00}(k_0)\Big] \Big[F^{0l0}(k_0)\Big]^*\Big\},
 \\ \nn
 \eqlab{K_l00_0l0}
  \calKKK_{l0}(\rrr) &= k_{l0}^\star\Big[\sin(k_{l0}^\star x)\cos(k_{l0}^\star y)\eee_x
  \\
  &\qquad \qquad -\cos(k_{l0}^\star x)\sin(k_{l0}^\star y) \eee_y\Big].
 \esubal
The impact on the bulk-driven streaming from the no-slip condition at the side walls will decrease exponentially with the length scale $w$ away from the side walls, where $w$ is the minimum shear length scale, \ie\, either the vertical length scale $\Lz$ or the horizontal length scale $(k_{l0}^\star)^{-1}$,
 \beq{solution_approximation}
 w = \min\big\{{\Lz,(k_{l0}^\star)^{-1}}\big\}.
 \eeq
For the narrow channel considered in \secref{3DSimulation} where $\Lz/\Lstar=0.1$, we have $w/\Lstar=0.1$ for $l=2$, $w/\Lstar=\frac{1}{4\pi}=0.08$ for $l=4$, and $w/\Lstar = \frac{1}{6\pi}=0.05$ for $l=6$, and therefore the vertical side walls can be ignored in the majority of the cavity for these three double modes.

To compute the acoustic streaming velocity, we ignore the side walls and assume that $(\vvv_{2}^\mr{blk})_{l0}$ is proportional to the $xy$-dependent acoustic energy flux density $\SSS_{l0}(x,y)$ and some $z$-dependent function $\zeta(z)$ to be found,
 \bal \eqlab{v2_bulk_l00_ansatz}
 (\vvv_{2}^\mr{blk})_{l0}=\kapfl\zeta(z)\SSS_{l0}(x,y).
 \eal
By using $\SSS_{l0}$ from \eqref{S_l00_0l0}, we find the three identities,
 \bsubal{lap_div_S_l00_0l0}
 \eqlab{divSl0}
 \div \SSS_{l0}&=0,
 \\
 \eqlab{LaplSl0}
 \lap \SSS_{l0}&=-2(k_{l0}^\star)^2\SSS_{l0},
 \\
 \eqlab{curlSl0}
 \curl \SSS_{l0}&= \frac{\big| \pa^{l00}\big|\big|\pa^{0l0} \big|\calF_{l0}}{\rhofl \omega} (k_{l0}^\star)^2
 \sin(k_{l0}^\star x)   \sin(k_{l0}^\star y)\een_z.
 \esubal
Consequently, by \eqref{divSl0}, the ansatz \eqnoref{v2_bulk_l00_ansatz} satisfies the continuity equation \eqnoref{v2_gov_cont_bulk}. Moreover, inserting~\eqnoref{v2_bulk_l00_ansatz} into \eqref{v2_gov_navier_bulk} using $\Gamfl=\Big(\frac43+\frac{\etaBfl}{\etafl}\Big)\etafl\kapfl\omega$ from \eqref{Gamfl_delta}, and \eqref{LaplSl0} we find,
 \bal
 \eqlab{gradp2S}
 \grad (p_{2}^{\mr{blk}})_{l0} &=
 \etafl\kapfl \bigg[\frac{\zeta''(z)}{(\sqrt{2}k_{l0}^\star)^2}-\zeta(z)+\frac12\bigg(\frac{4}{3}+\frac{\etaBfl}{\etafl}\bigg) \bigg]\SSS_{l0}.
 \eal
Since $\SSS_{l0}(x,y)$ has no $z$-component, $(p_{2}^{\mr{blk}})_{l0}$ on the left-hand side is independent of $z$. Therefore, the square bracket on the right-hand side must be a constant. This constant must be zero, since otherwise taking the curl of \eqref{gradp2S} would lead to $\zerovec=\curl \SSS_{l0}$ in conflict with \eqref{curlSl0}. The condition of a vanishing square bracket in \eqref{gradp2S} and the no-slip boundary conditions $\zeta(0)=\zeta(\Lz)=0$, lead to the following expression for $\zeta(z)$,
\beq{zeta_solution}
 \zeta(z)=\frac{1}{2} \bigg(\frac{4}{3}+\frac{\etaBfl}{\etafl}\bigg)\bigg[1-\frac{\cosh\big(\sqrt{2} k_{l0}^\star(z-\frac{\Lz}{2})\big)}{\cosh\big(\sqrt{2}k_{l0}^\star\frac{\Lz}{2}\big)}\bigg].
 \eeq
The solution for $(\vvv_{2}^\mr{blk})_{l0}$ is then obtained by inserting \eqsref{SRF_l00_0l0}{zeta_solution} in \eqref{v2_bulk_l00_ansatz}.

%\bibliographystyle{apsrev4-1-titles}
%\bibliography{acoustofluidics}
%\end{document}

%merlin.mbs apsrev4-1.bst 2010-07-25 4.21a (PWD, AO, DPC) hacked
%Control: key (0)
%Control: author (72) initials jnrlst
%Control: editor formatted (1) identically to author
%Control: production of article title (1) required
%Control: page (0) single
%Control: year (1) truncated
%Control: production of eprint (0) enabled
%

\end{document}